\documentclass[a4paper, 12pt]{article}

\usepackage[pages=all, color=black, position={current page.south}, placement=bottom, scale=1, opacity=1, vshift=5mm]{background}
\SetBgContents{
	\tt  
}      

\usepackage[margin=1in]{geometry} 

\usepackage{amsmath}
\usepackage{amsthm}
\usepackage{amssymb}

\usepackage[utf8]{inputenc}

\RequirePackage[colorlinks,citecolor=blue,urlcolor=blue]{hyperref}
\hypersetup{
	unicode,
	pdfauthor={Ziyi Song},
	pdftitle={Clustering Computer Mouse Tracking Data with Informed
Hierarchical Shrinkage Partition Priors},
	pdfsubject={A simple article template},
	pdfkeywords={article, template, simple},
	pdfproducer={LaTeX},
	pdfcreator={pdflatex}
}

\theoremstyle{plain}

\theoremstyle{definition}

\usepackage{graphicx, color}
\graphicspath{{fig/}}
\usepackage{subfigure}
\usepackage{enumitem}

\usepackage{algorithm, algpseudocode} 
\usepackage{mathrsfs} 

\usepackage{lipsum}

\RequirePackage{amsthm,amsmath,amsfonts,amssymb, soul}
\RequirePackage[authoryear, round]{natbib}

\RequirePackage[colorlinks,citecolor=blue,urlcolor=blue]{hyperref}
\RequirePackage{graphicx}

\usepackage{amsmath}        
\usepackage{bbm}            
\usepackage{amsfonts}       
\usepackage{nicefrac}       
\usepackage[many]{tcolorbox}
\newtcolorbox[]{algorithm1}{breakable, enhanced jigsaw, rounded corners, parbox=false, sharp corners, colback=white, colbacktitle=white,coltitle=black}

\usepackage{array}
\usepackage{adjustbox}
\usepackage{tikz}
\usepackage{tikzsymbols}
\usetikzlibrary{arrows, arrows.meta, automata, backgrounds, bayesnet, fit, positioning, calc}
\tikzset{
    state/.style={
           rectangle,
           rounded corners,
           draw=black, thick,
           minimum height=2em,
           inner sep=10pt,
           text centered,
           },
}

\usepackage{hyperref}
\hypersetup{
    colorlinks,
    citecolor=blue,
    filecolor=blue,
    linkcolor=black,
    urlcolor=black,
}
\usepackage{url}

\usepackage{xcolor}        
\usepackage{color}

\definecolor{lightyellow}{HTML}{ffffa7}
\definecolor{col_plate_R}{HTML}{61c1cf}
\definecolor{col_plate_Z}{HTML}{eb4702}
\definecolor{col_plate_N}{HTML}{02eb12}
\definecolor{col_plate_mu}{HTML}{b547e6}
\usepackage{bbm}

\usepackage{subcaption}

\title{Clustering Computer Mouse Tracking Data with Informed Hierarchical Shrinkage Partition Priors}

\author{Ziyi Song$^{1,*}$ \and Weining Shen $^1$ \and Marina Vannucci$^2$ \and Alexandria Baldizon$^3$ \and Paul M. Cinciripini$^3$ \and Francesco Versace$^3$ \and Michele Guindani$^{4,**}$}

\date{
	$^1$Department of Statistics, University of California, Irvine \\ 
	$^2$Department of Statistics, Rice University\\ 
        $^3$Department of Behavioral Science, The University of Texas MD Anderson Cancer Center\\
 	$^4$Department of Biostatistics, University of California, Los Angeles\\ 
  [2ex]
        $^*$ corresponding author: \texttt{ziyis9@uci.edu}\\
        $^{**}$ corresponding author: \texttt{mguindani@ucla.edu}
}

\begin{document}
\maketitle
	
\begin{abstract}
Mouse-tracking data, which record computer mouse trajectories while participants perform an experimental task, provide valuable insights into subjects' underlying cognitive processes. Neuroscientists are interested in clustering the subjects' responses during computer mouse-tracking tasks to reveal patterns of individual decision-making behaviors and identify population subgroups with similar neurobehavioral responses. These data can be combined with neuroimaging data to provide additional information for personalized interventions. In this article, we develop a novel hierarchical shrinkage partition (HSP) prior for clustering summary statistics derived from the trajectories of mouse-tracking data.  The HSP model defines a subjects' cluster as a set of subjects that gives rise to more similar (rather than identical) nested partitions of the conditions. The proposed model can incorporate prior information about the partitioning of either subjects or conditions to facilitate clustering, and it allows for deviations of the nested partitions within each subject group.  These features distinguish the HSP model from other bi-clustering methods that typically create identical nested partitions of conditions within a subject group. Furthermore, it differs from existing nested clustering methods, which define clusters based on common parameters in the sampling model and identify subject groups by different distributions. We illustrate the unique features of the HSP model on a mouse tracking dataset from a pilot study and in simulation studies. Our results show the ability and effectiveness of the proposed exploratory framework in clustering and revealing possible different behavioral patterns across subject groups. 
		
\end{abstract}

\hypertarget{introduction}{%
\section{Introduction}\label{section:intro}}

As neuroscience continues to reveal the complexity of cognitive processes, there is a growing recognition that to gain a comprehensive understanding of the relationship between decision-making and behavior, data must be collected across multiple modalities \citep{Noroozi2020}. 
In recent years, computer mouse tracking has become a popular and cost-effective technique for capturing continuous behavioral data. This technique entails acquiring the trajectory of the computer mouse as people are asked to perform a computer task. 
The resulting data can provide insights into the subjects' underlying decision-making cognitive processes. Neuroscientists are interested in combining the information from these behavioral data with knowledge acquired from neuroimaging data to provide additional insights about the neurobiological underpinnings of behavior and possibly inform personalized interventions.

\begin{figure}[t]
\centering
\includegraphics[width=10cm, height=6cm]{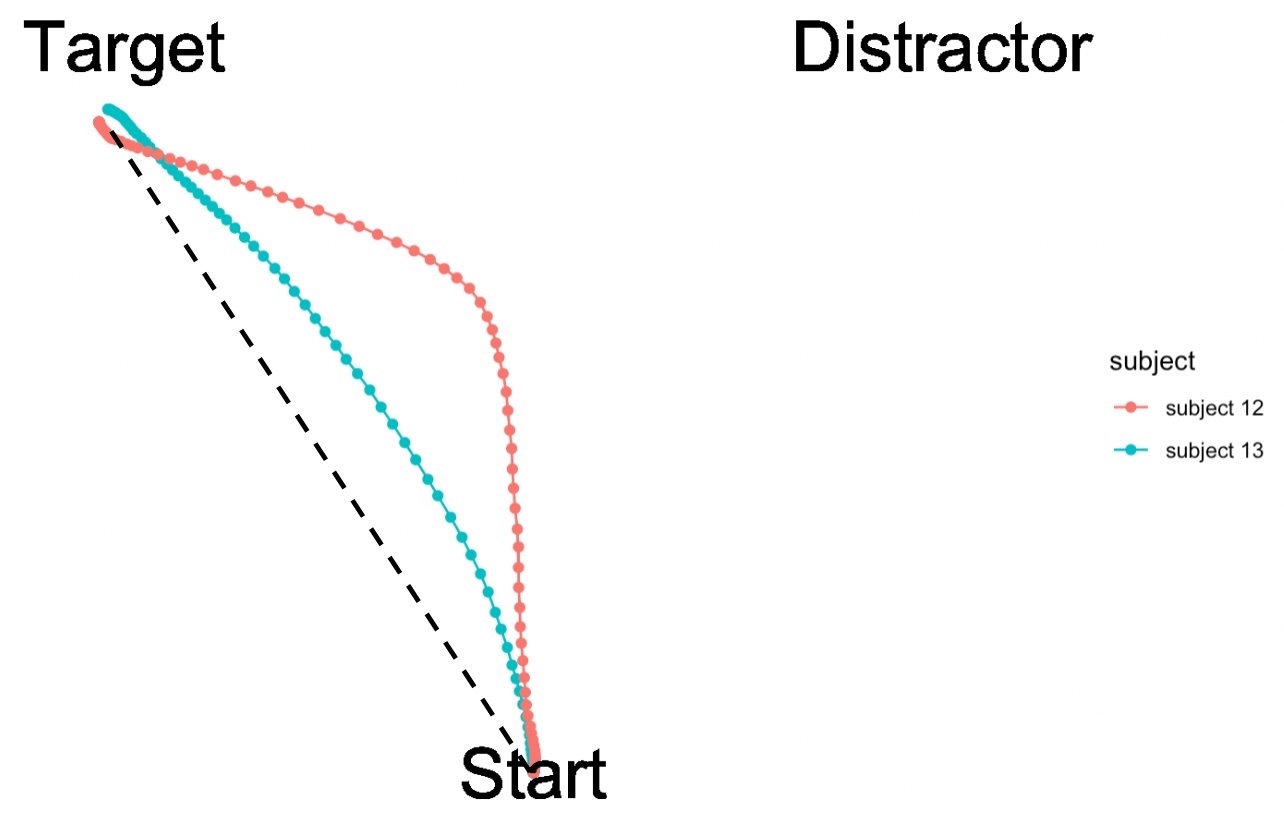}
\caption{An illustrative example of mouse-tracking data from the dataset considered in Section~\ref{Section: real.data.analysis}. The \emph{distractor} is a cigarette-related picture and the \emph{target} is a neutral picture. The black dashed line is the idealized straight line from the initial position of the mouse to the target. As an illustrative example, the red and the blue line are two trajectories drawn by two individuals, who have been grouped into two different subject groups based on the proposed HSP model. The MAD statistic captures the greatest point of deviation between the observed trajectories and the idealized straight line.}
\label{fig:trajectory.sample}
\end{figure}

In a typical mouse-tracking experiment designed to study attentional processes, participants are presented with a target and a distractor picture side-by-side on the screen and are instructed to click on the target. The underlying premise is that if the distractor captures the participant's attention, it will affect their behavior by deviating their hand trajectory toward the distractor. As the salience of the distractor increases, so does the magnitude of deviation away from the straight trajectory towards the target \citep{Dieciuc2019}. For example, individuals with substance use disorders (SUDs) tend to attribute high levels of motivational salience to drug-related cues. Hence, when a smoker is instructed to  click on a motivationally neutral picture, the presence of a cigarette picture can act as a distractor and cause the smoker's hand trajectory to deviate from the intended straight path towards the target image. The \emph{(target, distractor)} picture pairings form the experimental \emph{conditions}, which can be evaluated through multiple trials of the experiment.  See Figure~\ref{fig:trajectory.sample} for an illustration.

Measuring an individual's trajectory under different conditions can provide insight into their degree of response competition to alternatives and their decision-making processes.  The trajectories are typically summarized using simple statistics, e.g., the area under the curve (AUC) or the maximum absolute deviation (MAD). The AUC  measures  the area  between an idealized straight line going from the starting position of the mouse to the target, whereas MAD captures the greatest point of deviation from the idealized straight line \citep{Hehman2015}. 
Thus, data can be organized in a matrix with rows corresponding to values recorded under each condition and columns corresponding to subjects. 
Individual metrics are then grouped by condition, revealing patterns of behavior across picture pairs under the assumption of homogeneity in the subjects' responses. However, scientists are also interested in clustering subjects' responses within and between conditions to identify population subgroups with similar behaviors. 

In the Bayesian framework, nonparametric (BNP) approaches are typically employed for clustering subjects' responses into groups. Our application motivates a few considerations. First, scientists aim to classify subjects into distinct groups based on their behavioral responses to the presented choice pairs.  This suggests a \emph{nested clustering} of subjects and conditions within each cluster. There is a rich body of literature on nested clustering \citep[see, e.g.,][]{rodriguez2008nested, camerlenghi2019latent, beraha2021semi, denti2021common, lijoi2022flexible}. Those approaches typically allow for different distributions between subject groups, but within group data are still independent and identically distributed. As a different approach, \citet{lin2021separate} discussed a nested model by first defining groups of subjects and then nesting a partition of the conditions within each of the subject groups. As a result, all subjects in the same subject-group share the same nested partition among all the conditions. A similar approach has been previously proposed by \citet{lee2013nonparametric} for bi-clustering.

In the mouse tracking data application, however, the partitioning of the responses across conditions and clusters is typically more interesting than the actual values of the subjects' statistics. That is, the MAD values may vary across both trials and conditions, but they may still contribute to identifying the same subjects' groups (the same varying degrees of response competition across tasks). 
This leads us to consider random partition models, which directly define a probability distribution over alternative partitions \citep{Muller2010, muller2011product}. Also, subjects within a group can exhibit similar behavior, but this does not mean that their behavior must be necessarily identical across all conditions. When the number of choice pairs is large, it may be more reasonable to expect that a subject group is identified by partitions of the conditions that are just \emph{more similar} (rather than identical) within a group than between groups. 

Importantly, scientists might  have access to prior knowledge about the  possible partitioning of either subjects or conditions. For example, it is not uncommon for participants in a study 
 to undergo additional screenings, like completing surveys or engaging in other tests to assess their cognitive and behavioral performance. In the application discussed in Section~\ref{Section: real.data.analysis}, neuroscientists at the UT MD Anderson Cancer Center had previously used event-related potentials (ERPs, a direct measure of brain activity) to identify two groups of smokers who differed in their neuroaffective responses to pleasant cues and cigarette-related cues. Their neurobiological differences were associated to higher vulnerability to nicotine self-administration \citep{versace2023neuroaffective} and 
  greater risk of relapse during a smoking-cessation intervention  \citep{Versace2012-bb, Versace2014-az}. This phenomenon has been explained in terms of between group differences in the tendency to attribute incentive salience to drug-related cues, i.e., individuals who attribute higher incentive salience to nicotine-related cues than non-nicotine-related rewards are more prone to cue-induced compulsive behaviors \citep{Versace2023}. Neuroscientists are then interested in investigating to what extent the ERP results are confirmed when performing the (simpler and less expensive) mouse tracking experiment.

In this paper, we propose a hierarchical shrinkage partition (HSP) model to allow for bi-clustering of subjects and conditions within each subject group. The model further allows for small deviations of the individuals' partitions within each subject group, and for the incorporation of prior information into the model via the use of recently proposed shrinkage partition (SP) priors at the different levels of the hierarchy \citep{ dahl2023dependent}. 
While \citet{dahl2023dependent} suggest the use of their SP distribution for hierarchically dependent modeling of partitions, our HSP approach takes a targeted approach tailored to the specific needs of the neuro-behavioral application. It enables bi-clustering subjects and conditions, thereby introducing a new and distinct dimension to the framework. Compared to alternative approaches \citep[see, e.g.,][]{smith2019demand, paganin2021centered}, an SP prior is computationally efficient with increasing sample size and incorporates available knowledge by shrinking the partition probability function towards any \emph{a priori} base partition. In our model, we incorporate a base partition of the subjects informed by the results of an ERP experiment, and a base partition of the conditions based on elicited expert information about the expected population-level deviations from an idealized straight trajectory.

Many biclustering methods have appeared in the literature, beginning from the seminal papers by \citet{hartigan1972direct} and \citet{cheng2000biclustering}. Common models, such as the plaid model by \citet{lazzeroni2002plaid} and its refinement by \citet{turner2005improved}, focus on overlapping biclusters. That is, data are represented as a superposition of layers, each corresponding to a bicluster, thus allowing a data point to belong to multiple biclusters. While these methods originally assumed a predefined number of biclusters, the penalized prior approach of  \citet{chekouo2015penalized} and the truncated double stick-breaking prior of \citet{murua2022biclustering} have moved beyond this constraint, allowing for greater flexibility in determining the number of biclusters. Our model, motivated by our specific application, aims to group subjects with similar, nested partitions of the conditions. This results in distinct, non-overlapping submatrices, where each data point is associated exclusively with a single bicluster. This feature, akin to the methods of \citet{xu2013nonparametric} and \citet{lee2013nonparametric}, improves interpretability in our context. Furthermore, unlike typical biclustering that relies on shared parameter values for identifying submatrices, our model focuses on partition functions, thereby not depending on shared parameter values for linking the ERP and mouse-tracking data.

The remainder of this article is organized as follows. Section \ref{Section: Shrinkage Partition Prior} provides an introduction to a version of the shrinkage partition prior, first presented in \citet{dahlvideo}. Section \ref{Section: Method} describes the proposed HSP prior and investigates its clustering structure, which informs the posterior sampling algorithms outlined in the Section B of the Supplementary Material. An analysis of computer mouse tracking data is presented in Section \ref{Section: real.data.analysis}. A comparison of the empirical performance of our model with recent Bayesian nonparametric models on synthetic data is outlined in Section A of the Supplementary Material and succinctly highlighted in Section \ref{Section: simulation}. Finally, Section \ref{Section: discussion} contains concluding remarks and discussion of future work.

\hypertarget{sec:shrinkage-partition-distribution}{%
\section{Shrinkage Partition Distribution}\label{Section: Shrinkage Partition Prior}}

Shrinkage Partition distributions \citep[SP,][]{dahl2023dependent} can be seen as \emph{regularization priors} in Bayesian nonparametrics.  They modify the probability mass function of a Bayesian nonparametric model defined on the space of all partitions (\emph{baseline partition distribution}) to favor realizations that are close to an \emph{a priori} defined reference partition (\emph{base partition}), accommodating uncertainty and potential deviations from the initial partition. 

More formally, let $\boldsymbol{\pi}$ be a partition of the integers $\{1, \ldots, n\}$ into $S$ clusters, each of size $n_s$, $s=1,\ldots, S$. A partition  is defined by a set of cluster allocations $\boldsymbol{\pi} = (\pi_1, \ldots, \pi_n)$, with $\pi_j\in\{1, \ldots, S\}$, $j=1, 
\ldots, n$. Here, identical values of \( \pi_j \) indicate that the items belong to the same cluster. A widely employed distribution on the set of all partitions is the Chinese restaurant process (CRP), which is used to describe the partition structure of random variables distributed according to a Dirichlet Process \citep{Ferguson1973}. For any partition $\boldsymbol{\pi}$, a CRP($\beta$) assigns probability  $p_b(\boldsymbol\pi)=\left(\Gamma(\beta)/\Gamma(n+\beta)\right)\, \beta^S \, \prod_{s=1}^S \Gamma(n_s)$, with $\beta>0$ a mass parameter, characterizing the variance and clustering properties of the process. See, e.g., \citet{muller2015bayesian} for details. Alternative distributions can be employed as baseline, including the Pitman-Yor process \citep{PitmanYor1997}  and more general normalized completely random measures \citep{lijoi2010models}, provided their joint distribution can be computed through a sequential allocation scheme based on predictive probabilities. That is, $p_b(\boldsymbol\pi)$ can be computed as  $\prod_{t=1}^n \operatorname{Pr}_{b}\left(\pi_{t} \mid \pi_{1}, \ldots, \pi_{{t-1}}\right)$, with
$$
\operatorname{Pr}_{b}\left(\pi_{{t}} = s \mid \pi_{{1}}, \ldots, \pi_{{t-1}}\right) \propto\left\{
\begin{array}{ll}
\frac{\sum_{k=1}^{t-1}\, I\{\pi_{k}=s\} }{\beta+t-1}, \quad & s=1,\ldots,S^{(t)}, \\
\frac{\beta}{\beta+t-1}, \quad & s=S^{(t)}+1,
\end{array} \qquad t=1, \ldots, n,\right.
$$
where $S^{(t)}$ indicates the number of clusters within $\{\pi_1,\ldots, \pi_{t-1}\}$.
Notably, each term depends only on the cluster sizes; hence, for a CRP, the product is invariant to any permutation  $\boldsymbol{\delta} = (\delta_1, \ldots, \delta_n)$ of $\{1, \ldots, n\}$.  In the following, we employ the CRP as the baseline partition distribution to build a SP prior, following the formulation in \citet{dahlvideo}.

The \emph{a priori} information is encoded in a base partition, say $\boldsymbol{\nu}$. For a given permutation $\boldsymbol\delta$, the SP prior assigns the joint probabilities $p_{\mathrm{SP}}(\boldsymbol{\pi})$ through a sequential allocation scheme, 
\begin{align*}
p_{\mathrm{SP}}\left(\boldsymbol{\pi} \mid \boldsymbol{\nu}, \boldsymbol{\lambda}, \boldsymbol{\delta}, p_{\mathrm{b}}\right)  =  \prod_{t=1}^{n} \mathrm{Pr}_{\mathrm{SP}}\left(\pi_{\delta_t} \mid \pi_{\delta_1}, \ldots, \pi_{\delta_{t-1}}, \nu_{\delta_1}, \ldots, \nu_{\delta_t}, \lambda_t, p_{\mathrm{b}}\right),
\end{align*}
where $\boldsymbol{\lambda} = (\lambda_1, \ldots, \lambda_n)$ is a vector of non-negative shrinkage parameters that determine how close partitions drawn from the SP distribution are to the base partition $\boldsymbol{\nu}$. For any permutation $\boldsymbol{\delta}$,  $\mathrm{Pr}_{\mathrm{SP}}\left(\pi_{\delta_1} = 1 \mid \nu_{\delta_1}, \lambda, p_{\mathrm{b}}\right) = 1$, and
\begin{align}
\begin{split}
\mathrm{Pr}_{\mathrm{SP}}&(\pi_{\delta_t}=s \mid \pi_{\delta_1}, \ldots, \pi_{\delta_{t-1}},   \nu_{\delta_1}, \ldots, \nu_{\delta_t}, \lambda, p_{\mathrm{b}}) \\
&\propto
\begin{cases} 
\begin{array}{ll}
\mathrm{Pr}_{\mathrm{b}}\left(\pi_{\delta_t}=s \mid \pi_{\delta_1}, \ldots, \pi_{\delta_{t-1}}\right) \, \times \,
&
\exp \left(\lambda \;  \frac{\sum_{k=1}^{t-1} \mathrm{I}\left\{\pi_{\delta_k}=s\right\}\, \mathrm{I}\left\{\nu_{\delta_k}=\nu_{\delta_t}\right\}}{\sum_{k=1}^{t-1} \mathrm{I}\left\{\nu_{\delta_k}=\nu_{\delta_t}\right\}}\right)\\
& \mbox{if } s \in\left\{\pi_{\delta_1}, \ldots, \pi_{\delta_{t-1}}\right\}, \\
\mathrm{Pr}_{\mathrm{b}}\left(\pi_{\delta_t}=s \mid \pi_{\delta_1}, \ldots, \pi_{\delta_{t-1}}\right)  \, \times \, 
&
\exp \left(\lambda \; \mathrm{I}\left\{\sum_{k=1}^{t-1} \mathrm{I}\left\{\nu_{\delta_k}=\nu_{\delta_t}\right\}=0\right\}\right)\\
&  \mbox{if } s \notin\left\{\pi_{\delta_1}, \ldots, \pi_{\delta_{t-1}}\right\}, 
\end{array}
\end{cases}
\label{shrinkagepartitionconditionalfunction}
\end{split}
\end{align}
$t=1, \ldots, n$. Typically, $\lambda_t=\lambda$, $t=1, \ldots, n$. For large $\lambda$, the SP distribution favors partitions drawn from the CRP that are very similar to the base partition. This behavior is induced by the exponential term, which encourages alignment with the base partition. As $\lambda\to \infty$,  the SP ultimately converges to a point mass on the base partition $\boldsymbol{\nu}$. When $\lambda=0$, the SP distribution reduces to the CRP, i.e., $p_{\mathrm{SP}}(\boldsymbol{\pi})=p_{b}(\boldsymbol{\pi})$, independent of $\boldsymbol{\nu}$ and $\boldsymbol{\delta}$. However, in general, \eqref{shrinkagepartitionconditionalfunction} depends on $\boldsymbol{\delta}$. We treat the order of the permutation $\boldsymbol{\delta}$ as an additional parameter to estimate, with a uniform prior, $p(\boldsymbol{\delta}) = \frac{1}{n!}$. See also \citet{dahl2017random, dahl2023dependent}.  We indicate 
 a random partition $\boldsymbol{\pi}$ with SP distribution  as in \eqref{shrinkagepartitionconditionalfunction} by $\boldsymbol{\pi} \mid \boldsymbol{\nu}, \boldsymbol{\lambda}, \boldsymbol{\delta}, p_{\mathrm{b}}  \sim  
\mathrm{SP}\left(\boldsymbol{\pi} \mid \boldsymbol{\nu}, \boldsymbol{\lambda}, \boldsymbol{\delta}, p_{\mathrm{b}}\right)$.
\hypertarget{sec:HSP}{%
\section{Hierarchical Shrinkage Partition Prior}\label{Section: Method}}

We introduce our model in the context of the mouse-tracking data collected from $J$ subjects under $I$ conditions. Our data are represented as a $(I \times J)$ matrix, $\boldsymbol{Y}$, of continuous entries $y_{i,j}$'s, where each $y_{i,j}$ denotes a synthetic measure (e.g., MAD)  computed for the trajectory drawn by subject $j$ under condition $i$, for $i=1,\ldots,I$ and $j=1,\ldots,J$. Without loss of generality, we  assume $y_{i,j} \stackrel{\text{ind}}{\sim} f\left(\cdot \mid \theta_{i,j}\right)$, for some density function $f(y \mid \theta)$. In the following we assume a normal likelihood, $y_{i,j}\sim N(\mu_{i,j}, \sigma_{i,j}^{2})$ and $\theta_{i,j}=(\mu_{i,j}, \sigma_{i,j})$. Our first goal is to capture observed decision-making patterns by simultaneously partitioning the data across subjects and conditions. We begin by focusing on clustering conditions within each subject.

\subsection{Shrinkage Prior for clustering conditions}

Let $\boldsymbol{\pi}_{j}=\left\{\pi_{1, j}, \ldots, \pi_{I, j}\right\}$ now indicate a partition of the conditions $[I] = \left\{1, \ldots, I\right\}$ for each subject $j$ in $L_j \leq I$ clusters (\emph{condition grouping/partition}). 
Given a condition grouping $\boldsymbol{\pi}_{j}$, we assume that two conditions $i$ and $ i'$ assigned to the same cluster, i.e., such that $\pi_{i,j} = \pi_{i',j} = \ell$, are characterized by the same values of the $\theta$ parameters, i.e., $\theta_{i,j} = \theta_{i',j} = \theta_{\ell,j}^{\ast}$; hence, by a common prior $p\left(\theta^*_{\ell,j} \mid \boldsymbol{\pi}_{j}\right)$. The common values $\theta_{\ell,j}^{\ast}$'s are shared within  each subject $j$ but not across subjects. In the following, we assume a prior $H(\cdot)$ for the parameters $\theta_{\ell,j}^{\ast} = (\mu_{\ell,j}^{\ast}, \sigma_{\ell,j}^{\ast})$, as $\mu_{\ell,j}^{\ast} \stackrel{\text{iid}}\sim N(a_{j0}, b_{j0})$ and ${\sigma_{\ell,j}^{\ast 2}} \stackrel{\text{iid}}\sim \mathrm{Inverse-Gamma}(d_{j0}, e_{j0})$.

In a computer mouse-tracking experiment, conditions (i.e., the target-distractor pairings) are designed according to the \emph{a priori} expectations or hypotheses of the neuroscientists. Thus, integrating this knowledge is essential for making informed inferences about partitions. Specifically, let us assume a base partition, $\boldsymbol{\nu}_j=\{\nu_{1, j}, \ldots, \nu_{I, j}\}$, of the $I$ conditions, suggested by available expert information.
For example, in the context of our motivating application, we would like to analyze how the subjects' trajectories follow expected distraction patterns based on the specific types of targets and distractors presented. Then, following the discussion in Section \ref{Section: Shrinkage Partition Prior}, we assume a SP prior for any condition partition,   
\begin{equation}
    \boldsymbol{\pi}_j \mid \boldsymbol{\nu}_j, \boldsymbol{\lambda}, \boldsymbol{\delta}_j, \beta  \sim  
\mathrm{SP}\left(\boldsymbol{\pi}_j \mid \boldsymbol{\nu}_j, \boldsymbol{\lambda},  \boldsymbol{\delta}_j, \mathrm{CRP}(\beta) \right), 
\label{eq:SPpij}
\end{equation}
for a vector of shrinkage parameters, 
$\boldsymbol{\lambda}$, a permutation vector $\boldsymbol{\delta}_j$, $j=1, \ldots, J$, and a baseline partition distribution $p_{\mathrm{b}}=\mathrm{CRP}(\beta)$ with mass parameter $\beta > 0$.

\subsection{Hierarchical shrinkage for clustering subjects and conditions}

In the previous section, we employed an SP prior for clustering conditions within each subject. Notably, this clustering was independent across subjects. Thus, the base partitions had to be selected separately for each individual, a practice that is not viable in real-world scenarios. In this section, we tackle the issue of clustering subjects, as well as the hierarchical clustering of conditions across subjects, and their dependence on the availability of group- or population-level information.

Let $\boldsymbol{c}=\{c_1, \ldots, c_J\}$ denote a partition of the $J$ subjects into $|\boldsymbol{c}|$ clusters, where $|\cdot|$ denotes the cardinality of the partition. Expert information may also be available to inform the clustering of subjects. For instance, in the application discussed later in this paper, our neuroscientist collaborators had access to information about a grouping of subjects from an ERP experiment. One of their primary objectives was to determine whether the \emph{same} group of subjects exhibited discernible patterns when performing the mouse-tracking task. By incorporating this expert knowledge into our clustering analysis, we can potentially identify patterns and relationships across both the ERP and mouse-tracking datasets. We  also assume an SP prior to incorporate a base \emph{subject grouping}, say $\boldsymbol{c}_{0}$, i.e.,
\begin{equation}
    \boldsymbol{c} \mid \boldsymbol{c}_{0}, \boldsymbol{\tau}, \boldsymbol{\zeta}, \alpha_{0}
    \sim\operatorname{SP}\left(\boldsymbol{c} \mid \boldsymbol{c}_{0}, \boldsymbol{\tau}, \boldsymbol{\zeta}, \operatorname{CRP}\left(\alpha_{0}\right) \right),
\label{eq:SPc}
\end{equation}
where $\boldsymbol{\tau}$ is a non-negative shrinkage vector that reflects the degree of confidence neuroscientists place in the base subject grouping derived from the ERP experiment, $\boldsymbol{\zeta}$ is a permutation vector, and $\alpha_0$ is the mass parameter of a CRP process. Given a subject grouping $\boldsymbol{c}$, we anticipate that two subjects, say $j$ and $j'$, assigned to the same cluster will exhibit similar condition groupings. To encourage this, we posit that if $c_{j} = c_{j'} = k$, the corresponding base condition grouping will be identical as well, i.e., $\boldsymbol{\nu}_{j} = \boldsymbol{\nu}_{j'} = \boldsymbol{\nu}_{k}^{*}$, where $k\in\{1, \ldots, |\boldsymbol{c}|\}$ and $|\boldsymbol{c}| \leq J$. Thus, the individual base partitions $\boldsymbol{\nu}_{j}$ are characterized by the latent (shared) base partitions $\boldsymbol{\nu}_k^*$'s.  As a result, we anticipate that the two condition groupings, $\boldsymbol{\pi}_{j}$ and $\boldsymbol{\pi}_{j'}$, will be likely more similar,
although we flexibly allow for individual variations in a subject's decision-making behavior across conditions. Furthermore, we assume that the distinct partitions $\boldsymbol{\nu}_k^*$ are now influenced by a global base partition $\boldsymbol{\nu}_0$, which reflects the external neuroscience information. In the application discussed in Section \ref{Section: real.data.analysis}, the conditions can be classified into distinct categories \emph{a priori} based on shared distractors or shared targets. Then, the partitions $\{\boldsymbol{\nu}_{1}^{\ast}, \dots, \boldsymbol{\nu}_{|\boldsymbol{c}|}^{\ast}\}$ are assumed to be generated from $\boldsymbol{\nu}_{0}$ through a SP distribution as follows,
\begin{equation}
    \boldsymbol{\nu}_{k}^{\ast} \mid \boldsymbol{\nu}_{0}, \boldsymbol{\rho}, \boldsymbol{\epsilon}_{k}^{\ast}, \beta_{0}
    \overset{iid}{\sim} \operatorname{SP}\left(\boldsymbol{\nu}_{k}^{\ast} \mid \boldsymbol{\nu}_{0}, \boldsymbol{\rho}, \boldsymbol{\epsilon}_{k}^{\ast}, \operatorname{CRP}\left(\beta_{0}\right) \right), \quad k=1,\ldots,|\boldsymbol{c}|,
\label{eq:SPnukstar}
\end{equation}
where $\boldsymbol{\rho}$ is a non-negative vector indicating the degree of conviction that neuroscientists hold with regard to the base partition, and  $\boldsymbol{\epsilon}_{k}^{\ast}$ is a randomly generated permutation  of the conditions for the subject group $k$. Following the discussion in Section \ref{Section: Shrinkage Partition Prior}, as $\boldsymbol{\rho}\to\infty$, the partitions $\boldsymbol{\nu}_{k}^{\ast}$ align with the base partitions $\boldsymbol{\nu}_{0}$ with high probability.

We refer to the nested partition model described above as a Hierarchical Shrinkage Partition (HSP) prior. In summary, we have 
\begin{align}
\label{modelhsp}
\begin{split}
y_{i,j} \mid \pi_{i,j} = \ell &\overset{\text{ind}}{\sim} f\left(y_{i,j} \mid \theta_{\ell, j}^{\ast}\right),  \quad i=1, \ldots, I, \, j=1, \ldots, J,
\\
\theta_{\ell, j}^{\ast} &\overset{\text{iid}}{\sim} H, \quad \ell=1, \ldots, L_j, \, j=1, \ldots, J,\\
\boldsymbol{\pi}_{j} \mid \boldsymbol{\nu}_{j}, \boldsymbol{\lambda}, \boldsymbol{\delta}_j, \beta &\overset{\text{ind}}{\sim} \operatorname{SP}\left(\boldsymbol{\pi}_j \mid \boldsymbol{\nu}_{j}, \boldsymbol{\lambda}, \boldsymbol{\delta}_j, \operatorname{CRP}\left(\beta\right) \right),
\\
\boldsymbol{\nu}_{k}^{\ast} \mid \boldsymbol{\nu}_{0}, \boldsymbol{\rho}, \boldsymbol{\epsilon}_{k}^{\ast}, \beta_{0} &\overset{\text{iid}}{\sim} \operatorname{SP}\left(\boldsymbol{\nu}_{k}^{\ast} \mid \boldsymbol{\nu}_{0}, \boldsymbol{\rho}, \boldsymbol{\epsilon}_{k}^{\ast}, \operatorname{CRP}\left(\beta_{0}\right) \right),
\\
\boldsymbol{c} \mid \boldsymbol{c}_{0}, \boldsymbol{\tau}, \boldsymbol{\zeta}, \alpha_{0} &\sim\operatorname{SP}\left(\boldsymbol{c} \mid \boldsymbol{c}_{0}, \boldsymbol{\tau}, \boldsymbol{\zeta}, \operatorname{CRP}\left(\alpha_{0}\right) \right),
\end{split}
\end{align}
 with uniform priors assigned to the permutation parameters $\boldsymbol{\delta}_{j}$, $\boldsymbol{\epsilon}_{k}^{\ast}$, and $\boldsymbol{\zeta}$. Here, the shrinkage parameter $\boldsymbol{\lambda}$ determines the degree of similarity between the individual condition grouping $\boldsymbol{\pi}_{j}$ and the latent  condition  grouping $\boldsymbol{\nu}_{j}$, $j=1,\ldots, J$. Posterior inference is conducted via Markov Chain Monte Carlo (MCMC) methods. The algorithm is detailed in Section B of the Supplementary Material.

\hypertarget{sec:applications}{%
\section{Computer Mouse-tracking Data Analysis}\label{Section: real.data.analysis}}

\begin{figure}[H]
\centering
\includegraphics[width=\textwidth]{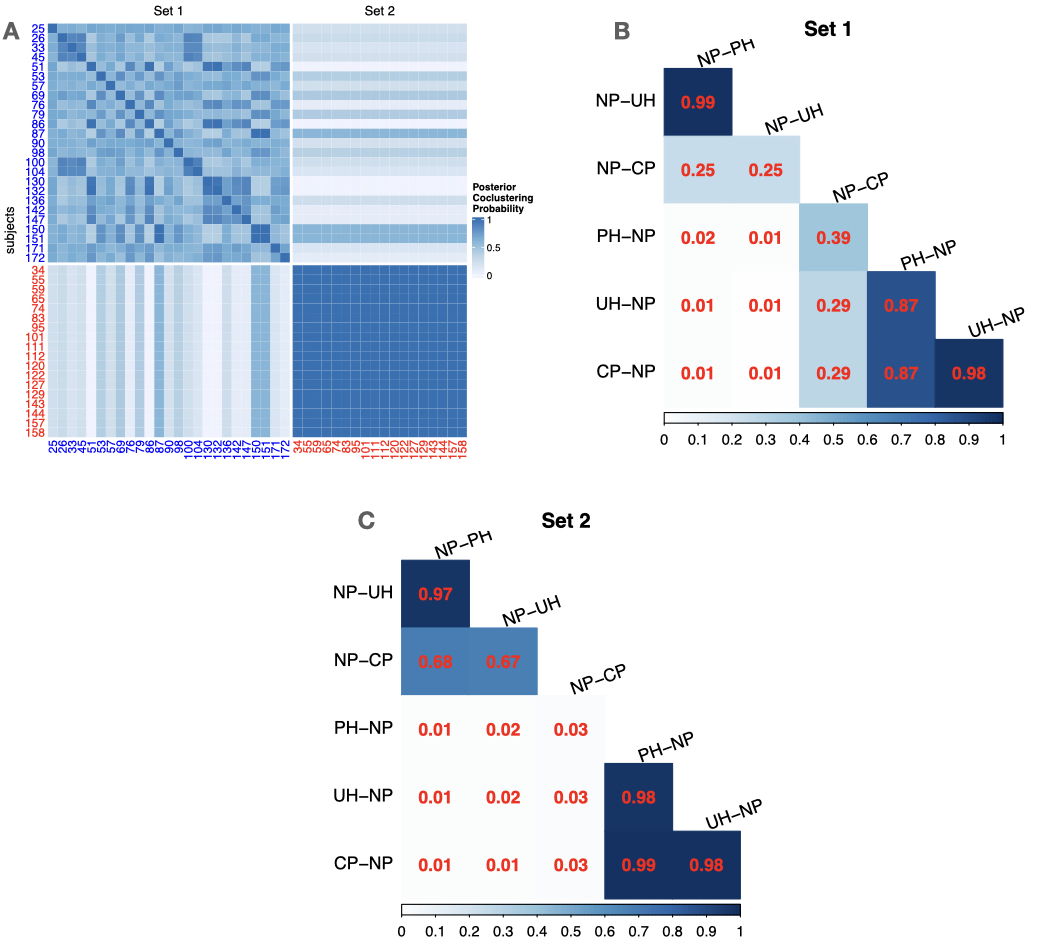}
\caption{\emph{Mouse tracking data}, depicting the clustering of subjects and mouse-tracking experimental conditions \emph{upon} constraining to the partition of subjects from previous ERP analyses ($\boldsymbol{\tau}=23, \boldsymbol{\rho}=4, \boldsymbol{\lambda}=20$). The top left panel displays the matrix of co-clustering probabilities for the subjects. The 43 subjects are color-coded into two groups, namely blue (ERP group 1) and red (ERP group 2). The two triangular matrices at the top right and the bottom report the co-clustering probabilities across conditions for ERP group 1 and ERP group 2, respectively. 
}
\label{fig:MAD.NP.23.4.20}
\end{figure}

We now consider the analysis of mouse-tracking data collected by our neuroscientist collaborators at the UT MD Anderson Cancer Center. The task was coded using E-prime2, a stimulus presentation software used in psychology experiments. The dataset comprises multiple trials resulting in mouse-tracking trajectories generated by 43 subjects (all daily smokers) across 6 experimental conditions.
The study's experimental conditions involved multiple picture pairings, with each consisting of a ``target" and a ``distractor" chosen from four categories of images: cigarette-related (CP), pleasant (PH), unpleasant (UH), and neutral (NP). To illustrate the conditions in a pair $(i,j)$, we denote the target followed by the distractor. For instance, the condition ``CP-NP" refers to a pairing where the target image is a CP picture and the distractor is an NP picture. For each subject, data were gathered from 16 trials under each condition. Each trial was time-normalized and summarized using the signed maximum absolute deviation (MAD) from the idealized straight line that connected the start point and the target picture. The MAD values from these trials were then averaged for each subject per condition. Consequently, this procedure yielded a total of 258 MAD values across 43 subjects and 6 conditions. Data were imported into the R package \emph{mousetrap} \citep{Kieslich2017} where specific movement tracking processing and analysis steps were completed according to \citet{wulff2021} to obtain the MAD for each condition and each subject. The choice of the MAD as a summary measure for mouse-tracking data was determined by its ability to capture critical regions of trajectories effectively, highlighting areas of maximum distraction, compared to other metrics like the Area Under Curve (AUC) and Average Deviation (AD). While these latter measures provide a comprehensive view of a trajectory, they tend to overemphasize the start and end points. In contrast, MAD, focusing on the single most deviant point, is simpler to interpret and better aligns with the goal of categorizing experimental conditions based on the distractor's influence on the trajectories.

Our collaborators are interested in clustering the behavioral patterns of the subjects during the mouse-tracking experiment, building upon the previous findings from an ERP experiment. Specifically, the analysis of the ERP data identified two distinct sets of smokers based on their observed brain responses. Smokers who exhibit higher responses during nicotine-related than pleasant images are more susceptible to nicotine self-administration and maladaptive cue-induced behaviors compared to smokers with the opposing neurocognitive profile \citep{Versace2023}. 
Moreover, the neuroscientists anticipate that the subjects' trajectories will exhibit expected patterns of distraction, contingent upon the type of targets and distractors. Using the population-level deviations from an idealized straight trajectory as a reference, the six conditions were classified into three groups: \{CP-NP, PH-NP, UH-NP\}, \{NP-CP\}, and \{NP-PH, NP-UH\}. These categories correspond to negligible deviations from a straight line, moderate deviations, and large deviations, respectively. These preliminary clusters of conditions and subjects function serve as the base partitions of conditions and subjects in the HSP model. They offer useful supplementary information beyond the mouse-tracking data to facilitate the hierarchical clustering process. It is worth noting that the two alternative methods we compare with in Section \ref{Section: simulation}, when applied to this data, classified all 43 subjects into a single group.

We apply the HSP model~(\ref{modelhsp}) to the matrix of data $\mathbf{Y}=\{y_{i,j}\}$ that includes $J=43$ subjects (column) and $I=6$ conditions (rows), standardized to have zero mean and unit variance across the conditions within each subject.  
If we reorganize the columns of the matrix according to the two groups of subjects identified by the ERP experiment, and align the matrix rows with the base groups previously outlined, the HSP model assumes the base partition of subjects as $\boldsymbol{c}_{0} = \{\boldsymbol{1}_{\bf 25},\boldsymbol{2}_{\bf 18}\}$, wherein 25 subjects are labeled 1 and 18 subjects are labeled 2, and the base partition of conditions as $\boldsymbol{\nu}_{0} = \{1,1,1,2,3,3\}$. 
The ERP findings suggest that the initial 25 subjects display a higher level of sensitivity towards non-cigarette-related rewards as opposed to cigarette-related cues. Conversely, the subsequent 18 participants exhibit a heightened sensitivity to cigarette-related cues. To differentiate between these groups, we will refer to them as ERP group 1 and ERP group 2, respectively.

We first investigate the behavioral patterns of subjects by strongly constraining the inference through shrinking the estimated partitions towards the ERP groups. We use the heuristic grid-search method described in Section D of the Supplementary Material to explore the clustering's sensitivity to several parameter values. The method evaluates the similarity between estimated partitions and the base groupings for conditions or subjects, using ARI and F1 measures.  Based on these sensitivity assessments, we set \(\boldsymbol{\tau} = 23\), \(\boldsymbol{\rho} = 4\), and \(\boldsymbol{\lambda} = 20\), to effectively integrate the domain information $\boldsymbol{\nu}_{0}$ and $\boldsymbol{c}_{0}$.
Figure~\ref{fig:MAD.NP.23.4.20} displays the results of this analysis. As expected, NP-PH co-clusters with NP-UH in both subject groups, since PH and UH serve as emotional distractors, and both subject sets display normal responses and are distracted by emotional images. Similarly, PH-NP and UH-NP are co-clustered in both sets, which is the counterpart to the co-clustering of NP-PH and NP-UH. These findings suggest that emotional distractors capture the subjects’ attentional resources in the same way, and they are not affected by the neutral distractors. Additionally, CP-NP co-clusters with PH-NP and UH-NP in both sets, which is consistent with the notion that neutral distractors do not distract subjects from salient images, being them cigarettes or emotional pictures. A noteworthy point pertains to the behavior of the condition NP-CP in the two groups. In group 1, NP-CP has a higher probability of co-clustering with PH-NP, UH-NP, and CP-NP compared to group 2. Thus, the subjects in group 1 do not appear to be significantly distracted by cigarettes-related pictures, similar to when neutral images are presented to them. This finding is consistent with the ERP results, indicating that these smokers do not attribute high incentive salience to  cigarettes-related cues. In contrast, NP-CP co-clusters with NP-PH and NP-UH in group 2. This suggests that smokers in group 2 continue to exhibit a response to cigarettes-related cues, consistent with the findings of the ERP experiments.

We can investigate the effect of partially utilizing or removing the ERP available information by selecting smaller values of the parameter $\boldsymbol{\tau}$. Here, we focus on the 
case where no ERP information is taken into account, by reducing $\boldsymbol{\tau}$ to 0, while keeping $\boldsymbol{\rho}$ = 4, $\boldsymbol{\lambda}$ = 20. 

\begin{figure}[H]
\centering
\includegraphics[width=\textwidth]{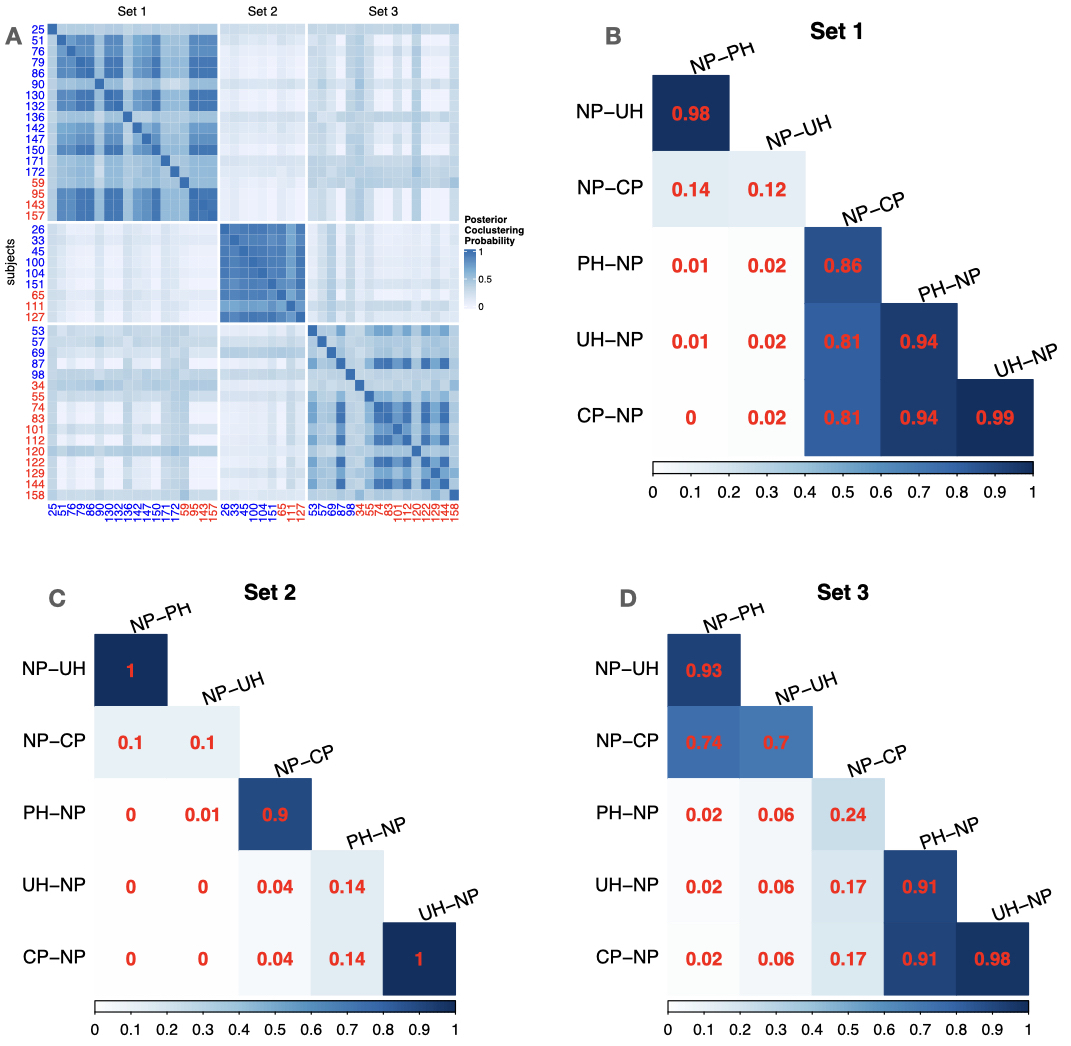}
\caption{\emph{Mouse tracking data}, depicting the clustering of subjects and mouse-tracking experimental conditions \emph{without} constraining to the partition of subjects from previous ERP analyses ($\boldsymbol{\tau}=0, \boldsymbol{\rho}=4, \boldsymbol{\lambda}=20$). The top left panel displays the matrix of co-clustering probabilities for the subjects. The 43 subjects are color-coded into two groups, namely blue (ERP group 1) and red (ERP group 2). The three triangular matrices report the co-clustering probabilities across conditions for the subjects' groups identified in our analysis. See Section \ref{Section: real.data.analysis} for more details.
}
\label{fig:MAD.NP.0.4.20}
\end{figure}

Figure~\ref{fig:MAD.NP.0.4.20} displays the results of this analysis. Notably, in addition to the two clusters identified by the previous analysis, a group of nine subjects (i.e., 26, 33, 45, 100, 104, 151, 65, 111, and 127), originally part of ERP group 1 and group 2, were identified as forming a distinct subset when ERP information was disregarded. Those subjects exhibit a different pattern of clustering for the conditions {NP-CP, PH-NP, UH-NP, CP-NP}, which are separated with high-probability into two distinct clusters, one consisting of {CP-NP, UH-NP} and the other consisting of {PH-NP, NP-CP}. This finding suggests that cigarettes as distractors behave similarly to neutrals for these smokers.

\hypertarget{sec:simulations}{%
\section{Simulation study}\label{Section: simulation}}

Section A of the Supplementary Material evaluates the performance of the proposed HSP through a simulation study in two scenarios and compares it with alternative methods. More specifically, the Section includes a comparison of our proposal with two alternative approaches: (1) the Nonparametric Bayesian Local Clustering (NoB-LoC) of \citet{lee2013nonparametric}; and (2) the Hidden Hierarchical Dirichlet Process (HHDP) of \citet{lijoi2022flexible}. The two methods capture different aspects of our approach. The former is a bi-clustering method; however,  a group of subjects is characterized by the same nested partition of the conditions. Our model, instead, is more flexible, as it partitions the conditions similarly, but not necessarily identically, across subjects assigned to a specific subject set (see Simulation 1 in Section A of the Supplementary Material).  Similarly to our approach, the HHDP  aims at clustering both subjects and conditions within subjects using a hierarchical model. It assumes that different groups of subjects are characterized by different \emph{distributions} of the conditions, instead of different nested partitions. As a result, different groups of subjects may have similar (but not necessarily equal) partitions of the conditions (see Simulation 2 in Section A of the Supplementary Material). In Section F of the Supplementary Material, we further compare our model's performance with the parametric Plaid biclustering model of  \citet{turner2005improved} under a scenario with a common condition grouping across all subjects.

\hypertarget{sec:discussion}{%
\section{Discussion}\label{Section: discussion}}

It is important to highlight the exploratory nature of our method, aiming to uncover different perspectives on the data. The role of the base partitions for rows and columns is critical, as it allows for the incorporation of valuable expert information that would be otherwise challenging to incorporate into an exploratory clustering procedure. At the same time, it is important to understand how much either partition should be trusted. For example, in considering a base partition of conditions (rows), it is also important to account for the degree of heterogeneity across all subjects in the original data. If we hold only a weak belief in the base partition of conditions, we are likely to observe many small clusters of subjects, while a stronger belief will likely lead to reduced heterogeneity and result in fewer, larger clusters of subjects. This can facilitate the clustering of columns,  yet it is important to balance the emphasis on the base partitions to avoid masking the true underlying patterns in the data. Hence, the importance of investigating the sensitivity to parameter selection. Although our proposal for such selection is heuristic, it is grounded in quantifying the similarity between the partition estimates and the base partitions for both conditions and subjects.

In the proposed HSP, the atoms that characterize the partitions of the conditions are not shared across subjects. That is, there is no notion of common atoms that allow for the clustering of rows across column clusters. Common atoms could be introduced into the framework, allowing the borrowing of information across all columns. Such borrowing of information could also help cluster prediction for new subjects, which is certainly of interest in applications, e.g., to inform new treatments.

Indeed, cluster prediction in biclustering is complex and multifaceted. Standard Bayesian methods could be used to predict a new subject's cluster membership by calculating posterior predictive probabilities based on the previously observed MAD values. Incorporating ERP classification information or covariates that influence behavioral clustering could make the prediction more targeted and help identify ERP-behavioral associations. 

Moving beyond Dirichlet processes (DPs) for the baseline partition distributions in the shrinkage processes is also a natural next step. Additionally, assigning priors to the vectors of shrinkage parameters to encourage the data to determine the degree of belief in the base partitions is another possible extension. However, it is essential to maintain computational feasibility and ensure the identifiability of all parameters based on the data.

Scalability is a challenge with many Bayesian nonparametric approaches. In particular, our algorithm, which is based on a modified Pólya urn scheme, requires calculating the partition probability function through a sequential allocation process. While faster alternatives exist, they may not adapt as effectively to the random partition model framework we use. Therefore, there is a need to develop more computationally efficient or approximate algorithms for inference. Potential alternatives might include split-merge moves or variational Bayes approaches.

While this paper focuses on modeling summary statistics from trajectory curves, directly modeling these curves could capture more complex patterns.  Functional data analysis techniques, such as functional principal component analysis (FPCA)  or functional regression, could be employed to simultaneously cluster subjects and conditions based on the shapes and patterns of the trajectory curves. For example, FPCA could be used to extract low-dimensional representations of the curves, capturing the main modes of variation across subjects and conditions. Moreover, extending the current framework to model the curves as vector-valued observations across conditions could account for correlations among different dimensions of the curves. This approach would allow for the discovery of more complex biclustering structures, where subjects and conditions are grouped based on multiple aspects of the trajectories, such as their velocity, acceleration, or curvature. All these options offer promising directions for future research.

\paragraph{Acknowledgements} The authors thank Dr. Giovanni Rebaudo and Dr. Juhee Lee for providing their codes for the comparisons with the HHDP and the NoB-LoC. This work was partially supported  by the National Institute on Drug Abuse of the National Institutes of Health (R01DA032581), by AACR Grant Number 19-90-52-CINC, and by MD Anderson’s Cancer Center Support Grant P30CA016672.  The content is solely the responsibility of the authors and does not necessarily represent the official views of the funding agencies that supported this work.

\paragraph{Supplementary Materials} Web Appendices and Figures referenced in Sections \ref{section:intro}, \ref{Section: Method}, \ref{Section: real.data.analysis}, and 5, along with code,  are available with this
paper at the Biometrics website on Oxford Academic. The code for our manuscript is also made available on GitHub
\texttt{https://github.com/Ziyi-Song-Stats/HSP.git}

\paragraph{Data Availability} Data is available upon reasonable request by contacting Francesco Versace at the following email address: \texttt{fversace@mdanderson.org}.
	
\bibliographystyle{plainnat} 
\bibliography{references-zotero.bib}       

\appendix

\newpage

\renewcommand{\thesection}{\Alph{section}}
\renewcommand{\thesubsection}{\Alph{section}.\arabic{subsection}}

\section{Simulation study}
\label{Section: simulation}

In this Section, we evaluate the performances of the proposed HSP with a simulation study comprised of two scenarios and perform comparisons with alternative methods. 
 Given knowledge of base partitions for either subjects or conditions, our goal is to assess the ability of the model to (i) identify subject groups that are more similar in terms of their condition groupings and (ii) cluster the conditions within each subject.  
 
More specifically, we compare our proposal with two alternative approaches: (1) the Nonparametric Bayesian Local Clustering (NoB-LoC) of \citet{lee2013nonparametric}; and (2) the Hidden Hierarchical Dirichlet Process (HHDP) of \citet{lijoi2022flexible}. The two methods capture different aspects of our approach. The former is a bi-clustering method; however,  a group of subjects is characterized by the same nested partition of the conditions. Our model, instead, is more flexible, as it partitions the conditions similarly, but not necessarily identically, across subjects assigned to a specific subject set (see Simulation 1).  Similarly to our approach, the HHDP  aims at clustering both subjects and conditions within subjects using a hierarchical model. It assumes that different groups of subjects are characterized by different \emph{distributions} of the conditions, instead of different nested partitions. As a result, different groups of subjects may have similar (but not necessarily equal) partitions of the conditions (see Simulation 2). In Section F of the Supplementary Material, we further compare our model's performance with the parametric Plaid biclustering model of  \citet{turner2005improved} under a scenario with a common condition grouping across all subjects.
  
 \emph{Simulation 1(a)} To compare with NoB-LoC, we simulated synthetic data comprising $J=60$ subjects tested on $I=30$ conditions, equally divided into three groups. The thirty data points in each subject were assumed to be partitioned differently across the three groups of subjects, with clustering labels $\{\boldsymbol{1_{5},1_{5},2_{5},2_{5},3_{5},3_{5}}\}$, $\{\boldsymbol{1_{5},2_{5},3_{5},1_{5},3_{5},2_{5}}\}$, and $\{\boldsymbol{1_{5},2_{5},1_{5},3_{5},2_{5},3_{5}}\}$, respectively, where $\boldsymbol{m}_{\bf n}$ denotes a vector of $n$ repetitions of the  label $m$ (e.g., $\boldsymbol{3}_{\bf 5} = \{3,3,3,3,3\} $). We generated $y_{ij} \sim N(\mu_{\ell,j}^{\ast}, \sigma_{\ell,j}^{2 \ast})$, with parameters chosen from estimating a three-component Normal mixture on the real data in Section 4 of the paper. For more details, see Section C of the Supplementary Material. Hence, 
 the locations $\mu_{\ell,j}^{\ast}$ of the condition clusters were generated by permuting $(-0.97,0.15,1.37)$ independently across all subjects, with $\sigma_{\ell,j}^{2 \ast} = 0.16$. We simulated 50 independent datasets for this scenario.
 
 Here, our aim is to illustrate the performance of our model without relying on any domain information derived from the base partition; hence, we set  $\boldsymbol{\tau} = \boldsymbol{\rho} = 0$ in model fitting. We selected $\boldsymbol{\lambda}=3.5$ based on a sensitivity analysis conducted over a range of $\boldsymbol{\lambda}$ values. This choice was informed by monitoring changes in the adjusted Rand Index (ARI) \citep[ARI,][]{rand1971objective, hubert1985comparing} and the symmetrized F1 measure  \citep[][]{murua2022biclustering} for the estimated partitions against the true values,  averaging across all subjects. For more details, refer to Section D in the Supplementary Material. We use a CRP prior for all the default partition priors $p_{\mathrm{b}}(\cdot)$, with $\alpha_{0} = \beta_{0} = \beta = 1$. For hyperparameters, we fix $a_{j0}$ and $b_{j0}$ at the sample mean and variance of $y_{ij}, i = 1, \ldots, I$, following  \citet{lee2013nonparametric} and \cite{lijoi2022flexible}. We specify $d_{j0} = 7.25$ and $e_{j0} = 1$ such that $E(\sigma_{\ell, j}^{2 \ast}) = 0.16$.

\begin{figure}[H]
\centering
\begin{subfigure}
\centering
\includegraphics[width=0.4\textwidth]{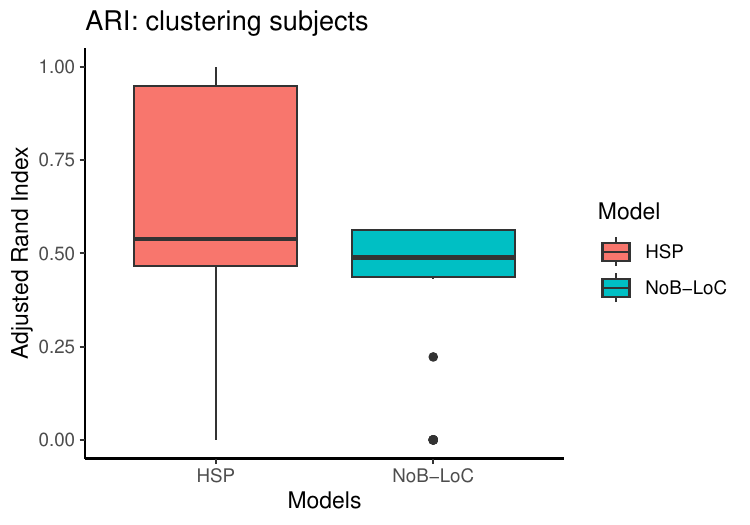}
\end{subfigure}
\hfill
\begin{subfigure}
\centering
\includegraphics[width=0.4\textwidth]{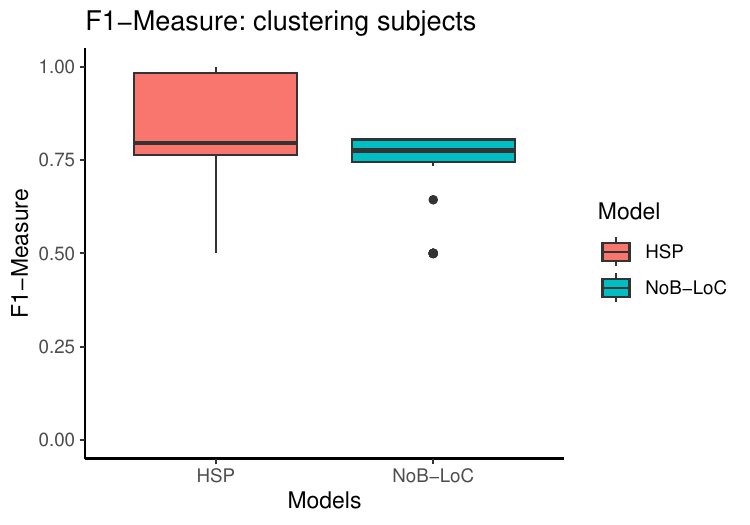}
\end{subfigure} 
\\
\begin{subfigure}
\centering
\includegraphics[width=0.4\textwidth]{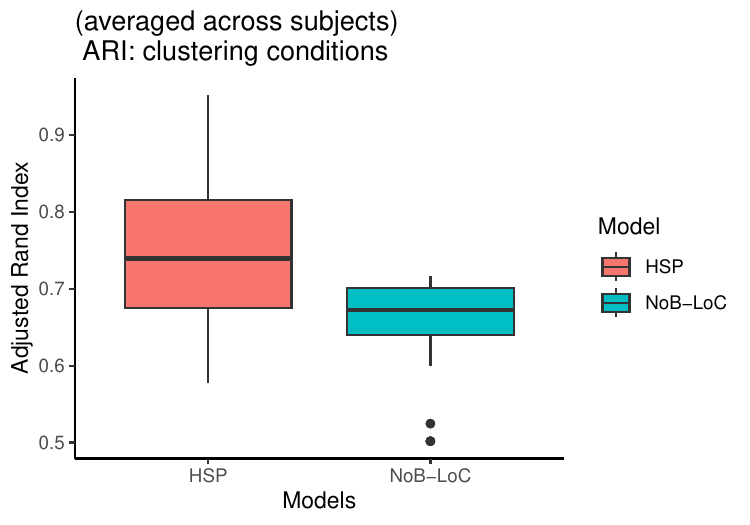}
\end{subfigure}
\hfill
\begin{subfigure}
\centering
\includegraphics[width=0.4\textwidth]{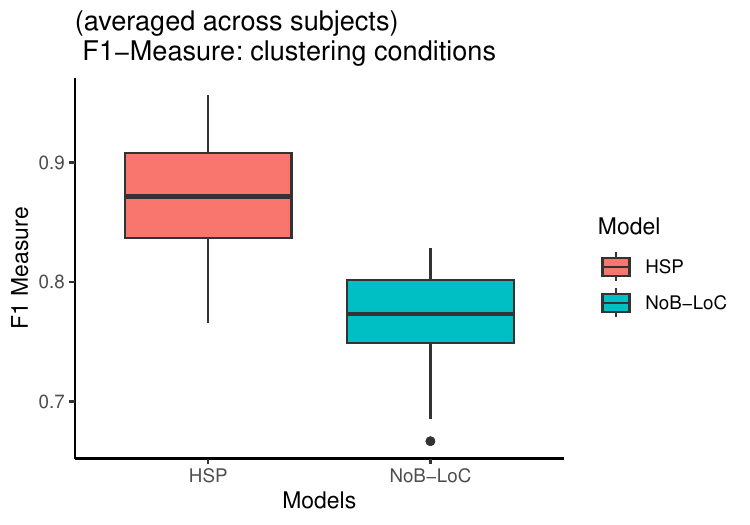}
\end{subfigure}
\caption{\emph{Simulation 1(a)}: Comparison between NoB-LoC and HSP across 50 replicates. HSP results were achieved with $\boldsymbol{\tau} = \boldsymbol{\rho} = 0$ and $\boldsymbol{\lambda} = 3.5$,  without leveraging any base partitions of subjects or conditions provided by prior knowledge. Partition estimations were evaluated using the adjusted Rand index (ARI) and symmetrized F1 measure, with closer to 1 indicating improved performance. The first row compares the partition of 60 subjects, while the second row compares the partition of 30 conditions. Both ARI and F1 measures were averaged over the subjects.}
\label{fig:Simulation.compare.HSP.NoBLoC}
\end{figure}

To perform posterior inference, we minimize the Variation of Information criterion \citep{wade2018bayesian} based on the MCMC output to estimate optimal partitions. See Figure~\ref{fig:Simulation.compare.HSP.NoBLoC}. Values of the adjusted Rand index and F1 measure nearing 1 indicate a close alignment between the identified structure and the true clustering. The HSP prior tends to perform better in the subject grouping and generally shows superiority in the condition grouping.  This improvement is partly due to the rigid bi-clustering structure of NoB-LoC: subjects might occasionally be inaccurately grouped, leading to a mismatch in their condition groupings. The more flexible hierarchical approach of the HSP helps to mitigate this issue. Section E of the Supplementary Material further illustrates how subject clustering is improved when increasing the number of subjects.

\begin{figure}[H]
\centering
\begin{subfigure}
\centering
\includegraphics[width=0.4\textwidth]{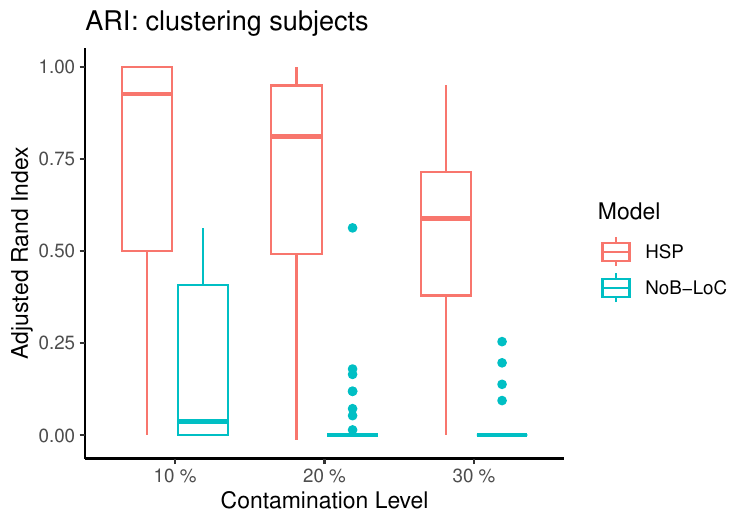}
\end{subfigure}
\hfill
\begin{subfigure}
\centering
\includegraphics[width=0.4\textwidth]{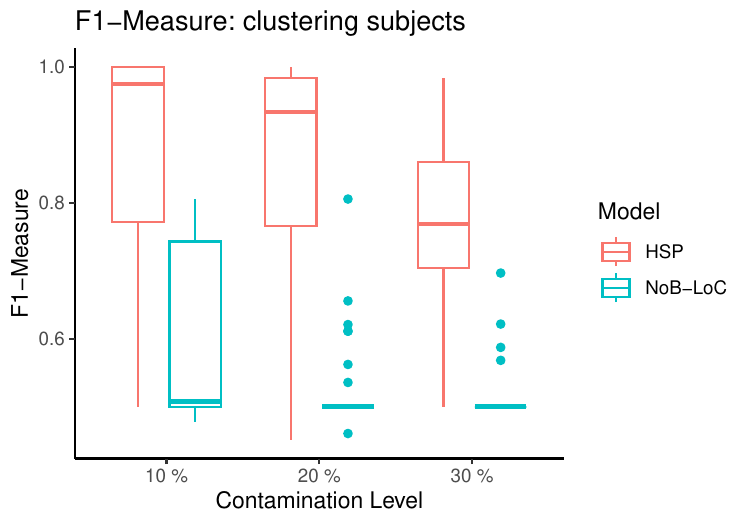}
\end{subfigure} 
\\
\begin{subfigure}
\centering
\includegraphics[width=0.4\textwidth]{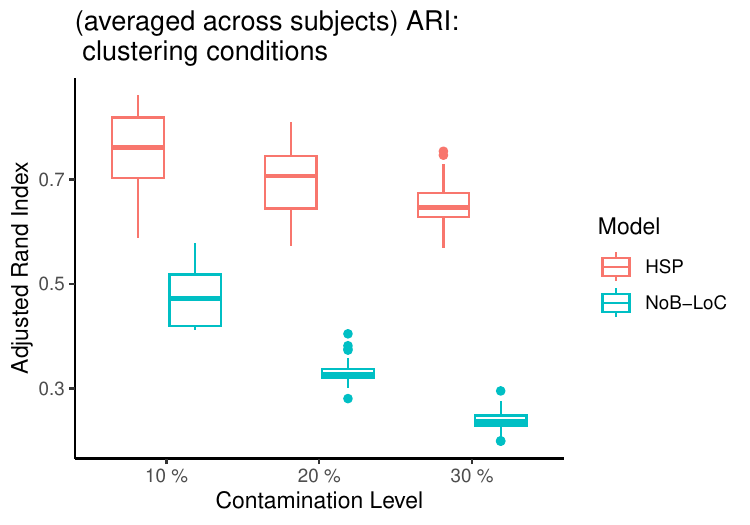}
\end{subfigure}
\hfill
\begin{subfigure}
\centering
\includegraphics[width=0.4\textwidth]{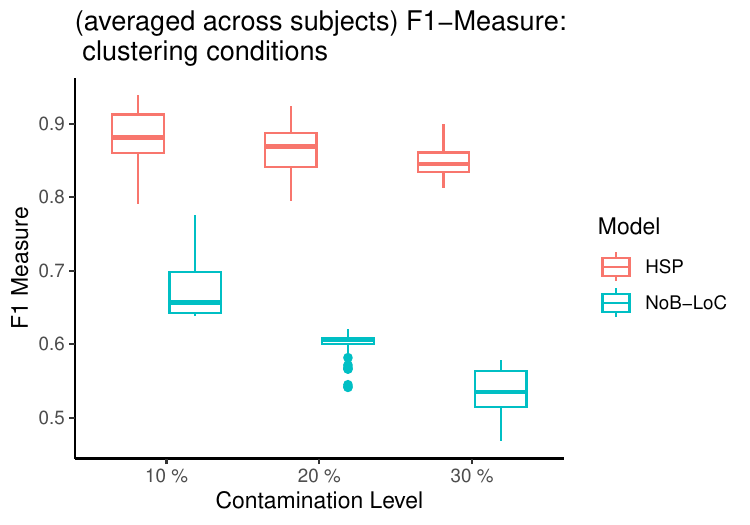}
\end{subfigure}
\caption{ \emph{Simulation 1(b)}: A comparison between the NoB-LoC and the proposed HSP on 50 replicates. The results for the HSP  were obtained by setting $\boldsymbol{\tau} = \boldsymbol{\rho} = 0$ and $\boldsymbol{\lambda} = 3.5$, that is, without leveraging any base grouping of subjects or conditions provided by prior knowledge. The performances of both NoB-LoC and HSP are further illustrated under increasing heterogeneity of the condition groupings within each subject group, a scenario referred to as contamination level ($10\%, 20\%, 30\%$)}.
\label{fig:Simulation.compare.HSP.NoBLoC.Setting1.2}
\end{figure}

\emph{Simulation 1(b)} We compare the HSP and the NoB-LoC also under a more challenging scenario where we contaminate the partition of the conditions in \emph{Simulation 1 (a)} for each subject. More specifically, for each subject's $I = 30$ cluster labels, we randomly select, say $10\%$,  of the labels and replace each of them with a new label chosen uniformly from $\{1,2,3\}$ to perturb the allocation patterns. This results in partitions of the conditions that are no longer completely identical across the subjects within a subject group, and we refer to this as a contamination level of $10\%$.  We then compare the ability of our HSP and NoB-LoC to cluster subjects and also conditions within each subject group, using a sequence of simulated data sets with increasing levels of contamination at $10\%, 20\%$, and $30\%$. We employ the same set up of Siulation 1(a) for model fitting. As shown in Figure~\ref{fig:Simulation.compare.HSP.NoBLoC.Setting1.2}, the HSP outperforms the NoB-LoC for clustering both subjects and conditions, especially when the partition patterns of the conditions are more heterogeneous within the sets of subjects. More in detail, both HSP and NoB-LoC naturally perform worse with increasing contamination levels. However, NoB-LoC tends to cluster all the subjects into one single group, resulting in a box-plot that shrinks to a constant with a few outliers. In contrast, the HSP tends to cluster all the subjects into smaller sets, providing at least some clustering information despite the less satisfactory performance on the ARI and F1 measures. As far as the condition grouping within each subject group, the HSP maintains a better performance for increasing contamination levels, while the performance of NoB-LoC drops dramatically. This is because NoB-LoC requires the subjects in a subject group to share fully identical partitions of the conditions, whereas the HSP allows for the existence of some heterogeneity in such partitions.

Notably, the results of Simulation 1 were obtained without employing any information about the base groupings of subjects or conditions. Section D of the Supplementary Material provides an example where $\boldsymbol{\tau}>0$, demonstrating its role in the estimation of the subject grouping.

\begin{figure}[H]
\centering
\begin{subfigure}
\centering
\includegraphics[width=0.45\textwidth]{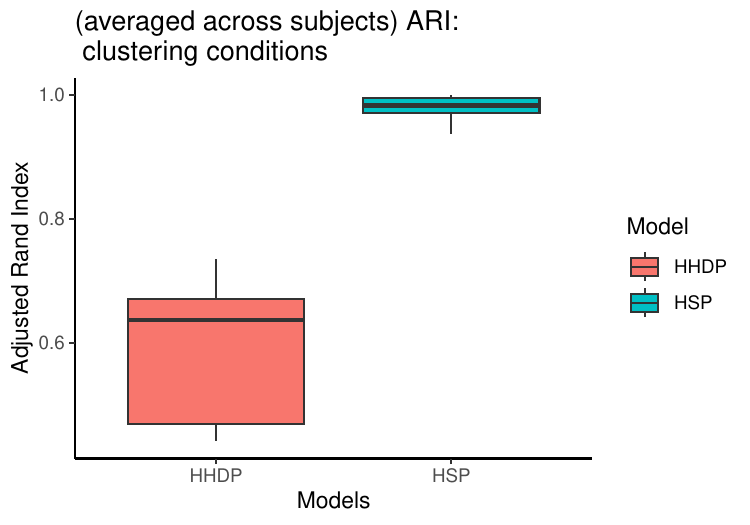}
\end{subfigure}
\hfill
\begin{subfigure}
\centering
\includegraphics[width=0.45\textwidth]{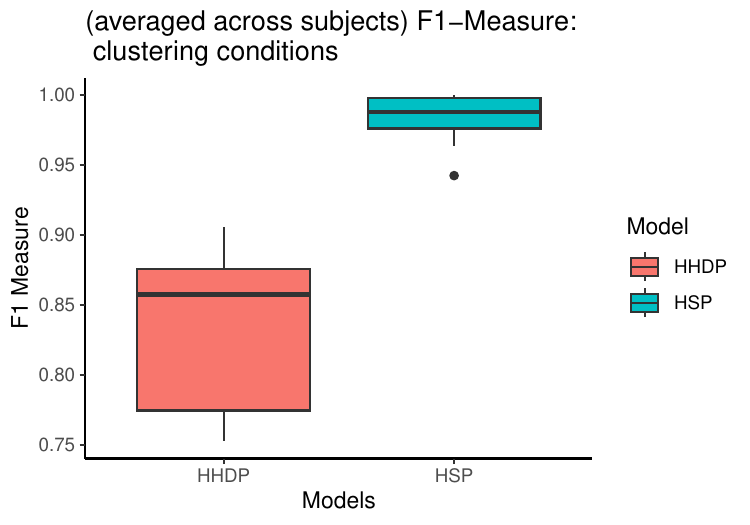}
\end{subfigure}
\caption{ \emph{Simulation 2:} A comparison between the HHDP and the proposed HSP on 50 replicates. The results of the HSP shown here are obtained assuming $\boldsymbol{\tau} = 0$ and $\boldsymbol{\rho} = 4, \boldsymbol{\lambda} = 3.5$} following the procedure detailed in Section D of the Supplementary Material.
\label{fig: simulation.compare.HHDP.HSP}
\end{figure}

\emph{Simulation 2}  We compare the HSP with the HHDP method of \citet{lijoi2022flexible}. Both frameworks allow bi-clustering of rows and columns in a data matrix, with the partitions of rows (conditions) being similar, but not necessarily equal,  within each column (subject) cluster. We employ the same true partitions as in \emph{Simulation 1(a)}. Following the scheme in \emph{Simulation 1(a)}, we generate locations $\mu_{\ell,j}^{\ast}$ of the condition clusters by permuting $(-0.97, 0.15, 1.37)$ independently across all the subjects, and then we generate data points $y_{ij} \sim N(\mu_{\ell,j}^{\ast}, \sigma_{\ell,j}^{2 \ast})$ with $\sigma_{\ell,j}^{2 \ast} = 0.16$. We incorporate a base partition of the conditions as $\boldsymbol{\nu}_{0} = \{\boldsymbol{1_5,2_5,3_5,4_5,5_5,6_5}\}$ to fit the HSP. As we are primarily interested in the effect of accounting for the base grouping of the conditions, we set again $\boldsymbol{\tau}=0$.  To select $\boldsymbol{\rho}$ and $\boldsymbol{\lambda}$, we use a step-by-step heuristic method, detailed in Section D of the Supplementary Material. This exploratory method evaluates the similarity between estimated partitions and the base groupings for conditions or subjects using ARI and F1 measures, investigating the clustering's sensitivity to various values of $\boldsymbol{\rho}$ and $\boldsymbol{\lambda}$. Based on such assessment,   we set $\boldsymbol{\rho} = 4$  and $\boldsymbol{\lambda}=3.5$ since these values seem to provide a balanced approach, aiming to maintain a reasonable level of similarity to the base partitions, without excessive influence (ARI and F1 measures close to 0.5) For the other hyperparameters, we use CRPs with total mass parameters $\alpha_{0} = \beta_{0} = \beta = 1$, $a_{j0}$ and $b_{j0}$ fixed at the sample mean and variance of $(y_{ij}, i=1,\ldots,I)$, and $d_{j0} = 7.25$ and $e_{j0} = 1$ such that $E(\sigma_{\ell,j}^{2 \ast}) = 0.16$. In the 50 simulation results, the HHDP clusters all the 60 subjects into one single group as we expect and the HSP correctly identifies the 3 true clusters of the subjects for most of the time. Nonetheless, we focus on the ability to cluster conditions within subjects in the present comparison. Figure~\ref{fig: simulation.compare.HHDP.HSP} shows that the proposed HSP performs better than the HHDP in identifying condition groupings. In this scenario, the HHDP clusters conditions into fewer groups within each subject, resulting in subject groups that cannot be distinguished due to similar partitions of conditions across all subjects.

The simulations were performed on a computing server equipped with Intel Xeon Gold 6240R processors, operating at 2.40 GHz with 192 GB of RAM. On average, our method took approximately 100 minutes to run 50 simulation replicates, each consisting of 10,000 MCMC iterations. We discarded the first 2000 iterations as burn-in.

\section{Posterior inference via MCMC}
\label{Section: posterior.inference}
Our Hierarchical Shrinkage Partition (HSP) prior is 
\begin{align*}
\label{modelhsp}
\begin{split}
y_{i,j} \mid \pi_{i,j} = \ell &\overset{\text{ind}}{\sim} f\left(y_{i,j} \mid \theta_{\ell, j}^{\ast}\right),  \quad i=1, \ldots, I, \, j=1, \ldots, J,
\\
\theta_{\ell, j}^{\ast} &\overset{\text{iid}}{\sim} H, \quad \ell=1, \ldots, L_j, \, j=1, \ldots, J,\\
\boldsymbol{\pi}_{j} \mid \boldsymbol{\nu}_{j}, \boldsymbol{\lambda}, \boldsymbol{\delta}_{j}, \beta &\overset{\text{ind}}{\sim} \operatorname{SP}\left(\boldsymbol{\pi}_j \mid \boldsymbol{\nu}_{j}, \boldsymbol{\lambda}, \boldsymbol{\delta}_{j}, \operatorname{CRP}\left(\beta\right) \right),
\\
\boldsymbol{\nu}_{k}^{\ast} \mid \boldsymbol{\nu}_{0}, \boldsymbol{\rho}, \boldsymbol{\epsilon}_{k}^{\ast}, \beta_{0} &\overset{\text{iid}}{\sim} \operatorname{SP}\left(\boldsymbol{\nu}_{k}^{\ast} \mid \boldsymbol{\nu}_{0}, \boldsymbol{\rho}, \boldsymbol{\epsilon}_{k}^{\ast}, \operatorname{CRP}\left(\beta_{0}\right) \right),
\\
\boldsymbol{c} \mid \boldsymbol{c}_{0}, \boldsymbol{\tau}, \boldsymbol{\zeta}, \alpha_{0} &\sim\operatorname{SP}\left(\boldsymbol{c} \mid \boldsymbol{c}_{0}, \boldsymbol{\tau}, \boldsymbol{\zeta}, \operatorname{CRP}\left(\alpha_{0}\right) \right),
\end{split}
\end{align*}
with uniform priors assigned to the permutation parameters $\boldsymbol{\delta}_{j}, \boldsymbol{\epsilon}_{k}^{\ast}$, and $\boldsymbol{\zeta}$.

To update the partitions $\left\{ \boldsymbol{c}, \boldsymbol{\nu}_{j}\left(j=1,\ldots,J\right), \boldsymbol{\pi}_{j}\left(j=1,\ldots,J\right) \right\}$ given the data $\boldsymbol{y}$ and all the other parameters, the marginal algorithms of \citet{neal2000markov} for updating a partition cannot be directly applied because the shrinkage partition (SP) distribution is non-exchangeable. However, since the p.m.f is available, it is possible to revise the algorithms by explicitly assuming partitions from the SP and by modifying the Gibbs samplers in \cite{neal2000markov} and \cite{dahl2017random}.

The parameters are estimated by iteratively drawing samples from the full conditional posterior distributions given the data and the other parameters:

\noindent
\renewcommand\labelenumi{\theenumi}
\begin{enumerate}[leftmargin=*]

\item Update $\theta_{\ell,j}^{\ast}$, $\ell = 1, \ldots, L_j; \, j = 1, \ldots, J$, using its full conditional distribution:
\begin{equation*}
P\left(\theta_{\ell,j}^{*} \mid y_{1,j}, \ldots, y_{I,j}, \ldots \right) \propto H\left(\theta_{\ell,j}^{*}\right) \times \prod_{i: \, \pi_{i,j} = \ell} f\left(y_{i,j} \mid \theta_{\ell,j}^{*}\right),
\end{equation*}
for $\ell = 1, \ldots, L_{j}$ and $j = 1, \ldots, J$. As described in the main text, here we assume a Gaussian kernel $f\left( \cdot \mid \theta_{\ell,j}^{\ast}\right)$ with $\theta_{\ell,j}^{\ast} = (\mu_{\ell,j}^{\ast}, \sigma_{\ell,j}^{2 \, \ast})$ and assume priors $\mu_{\ell,j}^{\ast} \sim \text{N}(a_{j0}, b_{j0})$ and $\sigma_{\ell,j}^{2 \ast} \sim \text{IG}(d_{j0}, e_{j0})$. Here $x \sim \text{IG}(a,b)$ indicates an inverse gamma distribution with $E(x) = b / (a-1)$.

\item Update the \emph{condition partition} $\boldsymbol{\pi}_{j}, j = 1, \ldots, J$.

For subject $j$, its \emph{condition partition} $\boldsymbol{\pi}_{j} = \{\pi_{1,j},\ldots,\pi_{I,j}\}$ can also be represented as a set of disjoint subsets $\{B_{1,j},\ldots,B_{L_j,j}\}$ with $|\boldsymbol{\pi}_{j}| = L_j$. Let $B_{1,j}^{-i}, \ldots, B_{L_j, j}^{-i}$ denote these subsets when condition $i$ is removed. Let $\boldsymbol{\pi}_{-i,j} = \{B_{1,j}^{-i}, \ldots, B_{L_j, j}^{-i}\}$ be the \emph{condition partition} with the label of condition $i$ removed. Furthermore, let $\boldsymbol{\pi}_{j}^{i \rightarrow \ell}$ be the \emph{condition partition} obtained by moving condition $i$ from its current subset to subset $B_{\ell, j}^{-i}$ with $\ell = 0,1,\ldots,L_j$. When $\ell = 0$, $\boldsymbol{\pi}_{j}^{i \rightarrow 0}$ indicates that condition $i$ is moved from its current subset to a newly created empty subset $B_{0,j}^{-i}$. We sequentially update a label $\pi_{i,j}$ for condition $i$ and subject $j$, $i = 1, \ldots, I$ and $j = 1, \ldots, J$ as follows:
\begin{equation*}
    p_{\mathrm{sp}}\left( i \in B_{\ell,j}^{-i} \mid \boldsymbol{\pi}_{-i,j}, \ldots \right) \propto p_{\mathrm{sp}}\left( \boldsymbol{\pi}_{j}^{i \rightarrow \ell} \mid \boldsymbol{\nu}_{j}, \boldsymbol{\lambda}, \boldsymbol{\delta}_{j}, \mathrm{CRP}\left(\beta\right)\right) \times f\left(y_{i,j} \mid \theta_{\ell,j}^{*}\right),
\end{equation*}
for $\ell = 0,1,\ldots,L_j$. When $\ell = 0$, $\theta_{0,\ell}^{\ast}$ is a new and independent sample from $H(\cdot)$, see Equation (5) of the main text. We update and increase the cardinality of the \emph{condition partition}    $\boldsymbol{\pi}_{j}$ by one if $\ell = 0$.

\item Update $\boldsymbol{\delta}_{j}$, the permutation of conditions for subject $j$.

We use Metropolis-Hastings algorithm to update $\boldsymbol{\delta}_{j}$. We propose a new permutation $\boldsymbol{\delta}_{j}^{\dagger}$ by randomly shuffling integers in the current permutation $\boldsymbol{\delta}_{j}$. As a symmetric proposal distribution, the proposed new $\boldsymbol{\delta}_{j}^{\dagger}$ is accepted with probability $$
\min\left\{\frac{p_{\mathrm{sp}}\left(\boldsymbol{\pi}_j \mid \boldsymbol{\nu}_{j}, \boldsymbol{\lambda}, \boldsymbol{\delta}_{j}^{\dagger}, \operatorname{CRP}\left(\beta\right)\right) p\left(\boldsymbol{\delta}_{j}^{\dagger}\right)}{p_{\mathrm{sp}}\left(\boldsymbol{\pi}_j \mid \boldsymbol{\nu}_{j}, \boldsymbol{\lambda}, \boldsymbol{\delta}_{j}, \operatorname{CRP}\left(\beta\right)\right) p\left(\boldsymbol{\delta}_{j}\right)}, 1\right\} = \min\left\{\frac{p_{\mathrm{sp}}\left(\boldsymbol{\pi}_j \mid \boldsymbol{\nu}_{j}, \boldsymbol{\lambda}, \boldsymbol{\delta}_{j}^{\dagger}, \operatorname{CRP}\left(\beta\right)\right)}{p_{\mathrm{sp}}\left(\boldsymbol{\pi}_j \mid \boldsymbol{\nu}_{j}, \boldsymbol{\lambda}, \boldsymbol{\delta}_{j}, \operatorname{CRP}\left(\beta\right)\right)}, 1\right\}
$$
since we are assuming a uniform distribution prior on the permutation.

\item Update the \emph{subject partition} $\boldsymbol{c} = \{c_1, \ldots, c_J\}$.

Following the notation in \cite{dahl2017random}, the current state of the partition $\boldsymbol{c} = \{c_1, \ldots, c_J\}$ is represented as a set of disjoint subsets $\left\{S_1, \ldots, S_K\right\}$. Let $S_{1}^{-j}, \ldots, S_{K}^{-j}$ denote these subsets when subject $j$ is removed. Correspondingly, let $\boldsymbol{c}_{-j}$ be the \emph{subject partition} with the label of subject $j$ removed. Furthermore, let $\boldsymbol{c}^{j \rightarrow k}$ be the \emph{subject partition} obtained by moving subject $j$ from its current subset to the subset $S_{k}^{-j}$, and let $\boldsymbol{c}^{j \rightarrow 0}$ denote the \emph{subject partition} obtained by moving subject $j$ from its current subset to an empty subset, say $S_{0}^{-j}$. We sequentially update a cluster label for subject $j$ as follows:
\begin{itemize}[leftmargin=*]
\item for an existing cluster label $k$ in $\boldsymbol{c}_{-j}$, subject $j$ is allocated to an existing cluster $k$ with probability,
\begin{equation*}
\begin{aligned}
    p_{\mathrm{sp}}(c_j = k \mid \boldsymbol{c}_{-j}, \ldots)  \propto p_{\mathrm{sp}}(\boldsymbol{c}^{j \rightarrow k} \mid \boldsymbol{c}_{0}, \boldsymbol{\tau}, \boldsymbol{\zeta}, \mathrm{CRP}(\alpha_0)) \\ \times
    p_{\mathrm{sp}}(\boldsymbol{\pi}_{j} \mid \boldsymbol{\nu}_{k}^{\ast}, \boldsymbol{\lambda}, \boldsymbol{\delta}_{j}, \mathrm{CRP}(\beta)),
\end{aligned}
\end{equation*}
where $\boldsymbol{\nu}_{k}^{\ast}$ is the current base partition shared by the subjects in the subset $S_{k}^{-j}$; 
\item subject $j$ is moved to a newly created empty subset, i.e. $k=0$, the probability is
\begin{equation*}
\begin{aligned}
    p_{\mathrm{sp}}(c_j = k \mid \boldsymbol{c}_{-j}, \ldots) \propto p_{\mathrm{sp}}(\boldsymbol{c}^{j \rightarrow k} \mid \boldsymbol{c}_{0}, \boldsymbol{\tau}, \boldsymbol{\zeta}, \mathrm{CRP}(\alpha_0)) \\ \times p_{\mathrm{sp}}(\boldsymbol{\pi}_{j} \mid \boldsymbol{\nu}^{new}, \boldsymbol{\lambda}, \boldsymbol{\delta}_j, \mathrm{CRP}(\beta)) ,
\end{aligned}
\end{equation*}
where $\boldsymbol{\nu}^{new}$ is a new base partition independently drawn from a $\operatorname{SP}\left(\cdot | \boldsymbol{\nu}_{0}, \boldsymbol{\rho}, \boldsymbol{\epsilon}_{k}^{\ast}, \operatorname{CRP}\left(\beta_{0}\right) \right)$, using Equation (4) of the main paper. We update the cardinality of the \emph{subject partition} $\boldsymbol{c}$ and the base partitions $\{\boldsymbol{\nu}_{1}^{\ast}, \ldots, \boldsymbol{\nu}_{|\boldsymbol{c}|}^{\ast}\}$
if subject $j$ is moved to a newly created empty subset. 
\end{itemize}

\item Update $\boldsymbol{\zeta}$, the permutation of subjects. 

Similarly, the proposed new permutation $\boldsymbol{\zeta}^{\dagger}$ is accepted with probability $$
\min\left\{\frac{p_{\mathrm{sp}}\left(\boldsymbol{c} \mid \boldsymbol{c}_{0}, \boldsymbol{\tau}, \boldsymbol{\zeta}^{\dagger}, \operatorname{CRP}\left(\alpha_{0}\right) \right)}{p_{\mathrm{sp}}\left(\boldsymbol{c} \mid \boldsymbol{c}_{0}, \boldsymbol{\tau}, \boldsymbol{\zeta}, \operatorname{CRP}\left(\alpha_{0}\right) \right)}, 1\right\}.
$$

\item Update the base partition $\left\{\boldsymbol{\nu}_1^{*}, \ldots, \boldsymbol{\nu}_{K}^{*}\right\}$, given the updated \emph{subject partition} $\boldsymbol{c} = \{c_1, \ldots, c_J\}$ with $|\boldsymbol{c}| = K$ from the previous steps.

A base partition $\boldsymbol{\nu}_{k}^{\ast} = \{\nu_{1,k}^{\ast}, \ldots, \nu_{I,k}^{\ast}\}$ can also be represented as a set of disjoint subsets $\{E_1, \ldots, E_D\}$. Let $E_1^{-i}, \ldots, E_{D}^{-i}$ denote these subsets with the $i$-th element $\nu_{i,k}^{\ast}$ removed. Let $\boldsymbol{\nu}_{-i,k}^{\ast}$ be the base partition $\boldsymbol{\nu}_{k}^{\ast}$ with the label $\nu_{i,k}^{\ast}$ removed. Let ${\boldsymbol{\nu}_{k}^{\ast}}^{i \rightarrow d}$ denote the base partition obtained by moving item $i$ from its current state $\nu_{i,k}^{\ast}$ to the subset $E_{d}^{-i}, d = 0,1,\ldots,D$. Here, $d=0$ indicates that the item $i$ is moved from its current state $\nu_{i,k}^{\ast}$ to a newly created empty subset. We sequentially update a label $\nu_{i,k}^{\ast}$, as $i=1,\ldots,I$ and $k=1,\ldots,K$ as follows:
\begin{align*}
    p_{\mathrm{sp}}(\nu_{i,k}^{\ast} = d \mid \boldsymbol{\nu}_{-i,k}^{\ast}, \text{ all the rest}) \propto \, & p_{\mathrm{sp}}({\boldsymbol{\nu}_{k}^{\ast}}^{i \rightarrow d} \mid \boldsymbol{\nu}_{0}, \boldsymbol{\rho}, \boldsymbol{\epsilon}_{k}^{\ast}, \mathrm{CRP}(\beta_0)) \\
    & \times 
    \prod_{j: \, c_{j} = k} p_{\mathrm{sp}}(\boldsymbol{\pi}_{j} \mid {\boldsymbol{\nu}_{k}^{\ast}}^{i \rightarrow d}, \boldsymbol{\lambda}, \boldsymbol{\delta}_{j}, \text{CRP}(\beta)),
\end{align*}
with $d = 0,1,\ldots,D$.

\item Update $\boldsymbol{\delta}_{k}^{\ast}$, the latent permutation of conditions for subject group $k$. 

We propose a new permutation $\boldsymbol{\epsilon}_{k}^{* \dagger}$ by randomly shuffling integers in the current permutation $\boldsymbol{\epsilon}_{k}^{*}$. As a symmetric proposal distribution in Metropolis-Hastings algorithm, the proposed new $\boldsymbol{\epsilon}_{k}^{* \dagger}$ is accepted with probability $$
\min\left\{\frac{p_{\mathrm{sp}}\left(\boldsymbol{\nu}_{k}^{\ast} \mid \boldsymbol{\nu}_{0}, \boldsymbol{\rho}, \boldsymbol{\epsilon}_{k}^{* \dagger}, \operatorname{CRP}\left(\beta_{0}\right) \right)}{p_{\mathrm{sp}}\left(\boldsymbol{\nu}_{k}^{\ast} \mid \boldsymbol{\nu}_{0}, \boldsymbol{\rho}, \boldsymbol{\epsilon}_{k}^{*}, \operatorname{CRP}\left(\beta_{0}\right) \right)} , 1\right\}.
$$   

\end{enumerate}

\vskip 1cm

\section{Simulations: determining the cluster locations for the data generating process}

\begin{figure}[H]
\centering
\begin{minipage}{.6\textwidth}
\centering
\includegraphics[width=\textwidth]{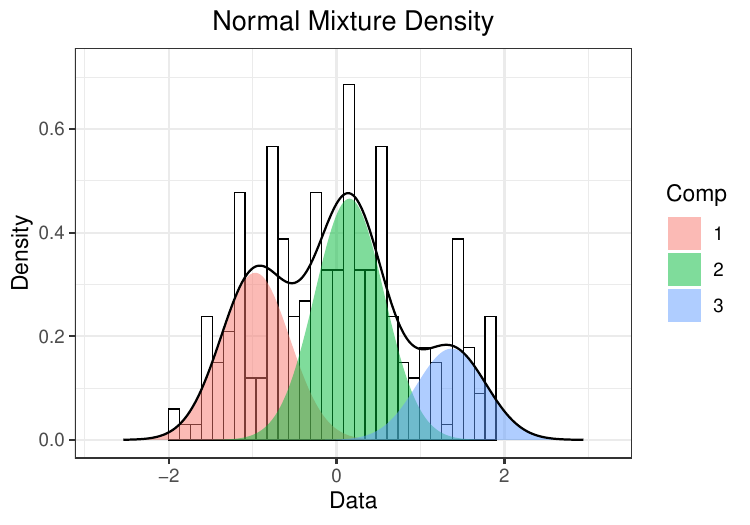}
\end{minipage}%
\caption{Means and standard deviations for the 3 components are: (-0.9699, 0.4138), (0.1532, 0.4138), and (1.3701, 0.4138).}
\label{fig:Reviewer1Point6_mimic_realdata}
\end{figure}

Our simulation setup has been defined to accurately replicate the cluster locations observed in the real data application. More specifically, we have employed the \emph{mixfit()} function from the R \emph{mixR} Package, to estimate a mixture model consisting of three Normal components. The resulting means and standard deviations of the three Normal components were (-0.9699, 0.4138), (0.1532, 0.4138), and (1.3701, 0.4138), as seen in Figure~\ref{fig:Reviewer1Point6_mimic_realdata}. 

\newpage

\section{Selection of shrinkage hyperparameters and sensitivity analysis}

The proposed HSP method is fundamentally exploratory, similar in approach to the K-means algorithm and other clustering algorithms. This means that our method does not deliver formal statistical inference, but it is meant to uncover different views of the data. Such insights are invaluable for hypothesis generation, facilitating the identification of novel patterns and relationships across modalities. 

Nevertheless, it is important to investigate the sensitivity of the clustering results to parameter selection. Here, we outline a heuristic procedure for such selection, grounded in quantifying the similarity between the partition estimates and the base partitions for both conditions and subjects. This evaluation involves calculating the Adjusted Rand Index (ARI) and the F1 measure over a grid of parameter values. Specifically, in the analysis of real data, we assess the estimated subject grouping against the base subject partition, and - similarly - we consider the estimated condition grouping compared to the base condition partition, averaged across all subjects. These measures provide some understanding of the model's dependency on base partition data. In particular, to determine the most appropriate parameter values, we closely monitor the ARI and F1 measures, focusing on points where they exhibit either a notable jump or achieve a state of relative stability, beyond which there are no significant increases. 

In the following, it is necessary to differentiate between Simulations 1(a) and 1(b) on one side, and Simulation 2 along with the data analysis on the other, owing to the differing objectives inherent in each analysis.

\vskip 2mm

{\bf Simulation 1.} Here, we seek to highlight the differences with the NoB-LoC method by \cite{lee2013nonparametric}; hence, the base partitions are not relevant and we set $\boldsymbol{\tau} = 0, \boldsymbol{\rho} = 0$. Indeed, we refrain from incorporating any existing domain information from the base partition. In this Simulation, our aim is to demonstrate that our approach is effective and can even outperform NoB-LoC without relying on any domain information derived from the base partition. We provide a proof-of-concept simulation of the case where $\boldsymbol{\tau}>0$ later in the present Section of the Supplementary material.

In Figure ~\ref{Fig: R1P7_select_lambda}, we illustrate the sensitivity of the proposed HSP model to variations in the parameter $\boldsymbol{\lambda}$, showcasing its impact on performance across the 50 generated datasets. As mentioned, we consider the ARI and F1 measure of the estimated subject partition and the ARI/F1 measure of the estimated condition partitions averaging across all subjects. Values of the ARI/F1 measure close to 1 indicate a close match with the truth. We employ a grid-like approach to vary the parameter $\boldsymbol{\lambda}$, considering values from 1 to 6 in increments of 0.5. For $\boldsymbol{\lambda}$ close to zero, the condition grouping $\boldsymbol{\pi}_j$ is independent of the latent base partition $\boldsymbol{\nu}_j$. In this scenario, we expect an increase in the variability, affecting both the subject grouping and the condition grouping. This is similar to the effect observed in Bayesian hierarchical models, where an unconstrained middle layer could induce excessive flexibility. 

Indeed, Figure~\ref{Fig: R1P7_select_lambda} shows that the performance of the subject and condition groupings can be affected by extreme values of $\boldsymbol{\lambda}$. In particular, high values of $\boldsymbol{\lambda}$ may also negatively impact the performance of the model when there's significant variation in condition groupings across different subject groupings, a scenario implied by our simulation setup. Based on the results of Figure ~\ref{Fig: R1P7_select_lambda}, we set $\boldsymbol{\lambda}=3.5$, a value that appears to balance an effective clustering of both subjects and conditions. In \emph{Simulation 1(b)}, our goal is to evaluate the model's performance in a more challenging scenario. We introduced contamination to the partition of conditions from \emph{Simulation 1(a)} for each subject. Consistently with \emph{Simulation 1(a)}, we maintained the parameters $\boldsymbol{\tau}$ and $\boldsymbol{\rho}$ at 0, and set $\boldsymbol{\lambda}$ at 3.5.

\begin{figure}[H]
\centering
\begin{subfigure}
\centering
\includegraphics[width=0.45\textwidth]{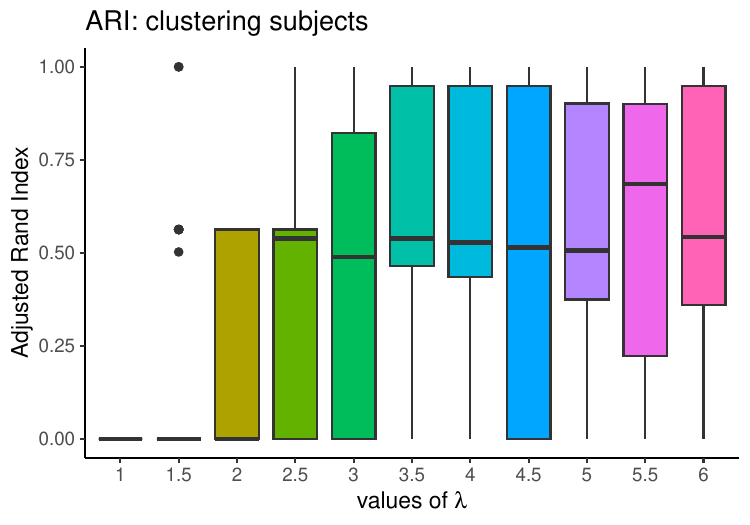}
\end{subfigure}
\begin{subfigure}
\centering
\includegraphics[width=0.45\textwidth]{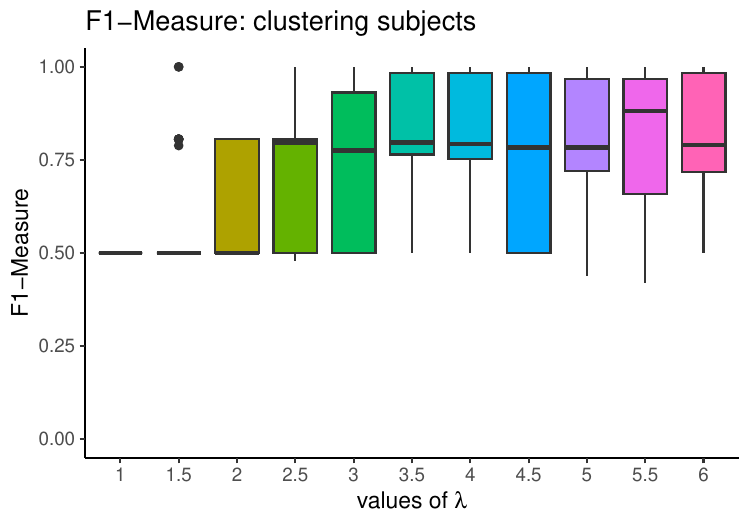}
\end{subfigure}
\\
\begin{subfigure}
\centering
\includegraphics[width=0.45\textwidth]{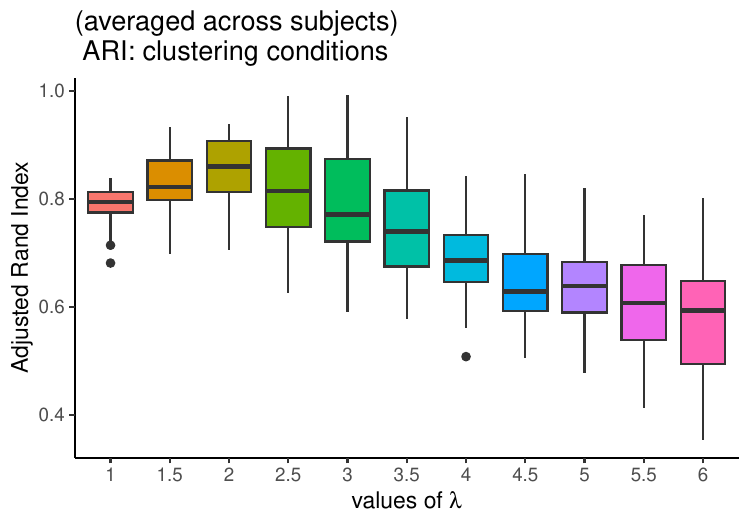}
\end{subfigure}
\begin{subfigure}
\centering
\includegraphics[width=0.45\textwidth]{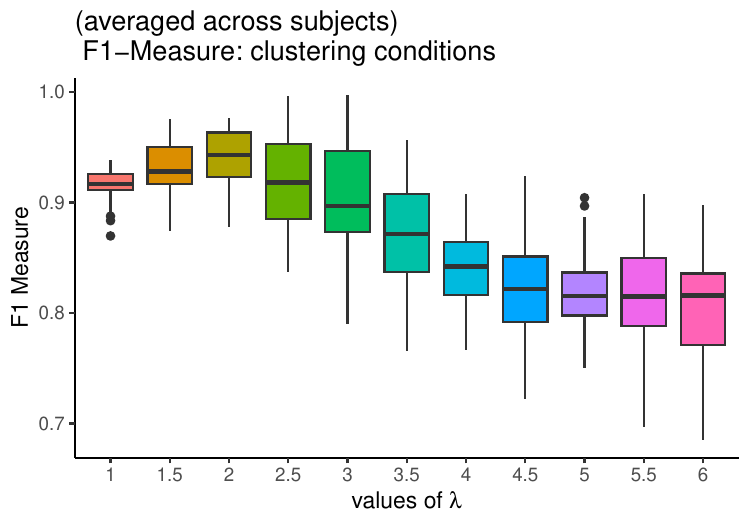}
\end{subfigure}
\caption{\emph{Simulation 1(a): Sensitivity analysis}. ARI and F1 measures on a grid of values of $\boldsymbol{\lambda}$ between 1 and 6 in increments of 0.5 ($\boldsymbol{\tau} = \boldsymbol{\rho} = 0$). The first row shows the performance for the partition of the 60 subjects; the second row shows the performance for the partition of the 30 conditions.}
\label{Fig: R1P7_select_lambda}
\end{figure}

{\bf{Simulation 2 and Data analysis.}} In both Simulation 2 and the Data analysis, we use a step-by-step heuristic method to select the parameters. This method assesses how closely the estimated partitions match the base partitions for conditions or subjects. The steps of the procedure can be described as follows:

\begin{enumerate}
    \item Start by setting $\boldsymbol{\tau} = 0$ (e.g., start by assuming the ERP subject grouping is not relevant).
    \item Define a grid of values for $\boldsymbol{\lambda}$ and $\boldsymbol{\rho}$.
    \item Select a random value for $\boldsymbol{\rho}$ from its range and compute the ARI and F1 measure for the $\boldsymbol{\lambda}$ values.
    \item Based on the inspection of these measures and the goal of the analysis, choose a value for $\boldsymbol{\lambda}$.
    \item With the selected $\boldsymbol{\lambda}$, reassess the ARI and F1 measures across the original range of $\boldsymbol{\rho}$ values.
    \item Based on the inspection of these measures, choose a value for $\boldsymbol{\rho}$.
    \item With the selected $\boldsymbol{\lambda}$ and $\boldsymbol{\rho}$, assess the ARI and F1 measures over a range of $\boldsymbol{\tau}$ values.
     \item Based on the inspection of these measures, choose a value for $\boldsymbol{\tau}$.
\end{enumerate}

The procedure offers the benefit of evaluating how the clustering is influenced by varying the parameters, effectively assessing the sensitivity to different parameter values.  In practice,  it's possible to bypass certain steps based on the objectives of the analysis. 

\vskip 2mm

\begin{figure}[H]
\centering
\begin{subfigure}
\centering
\includegraphics[width=0.35\textwidth]{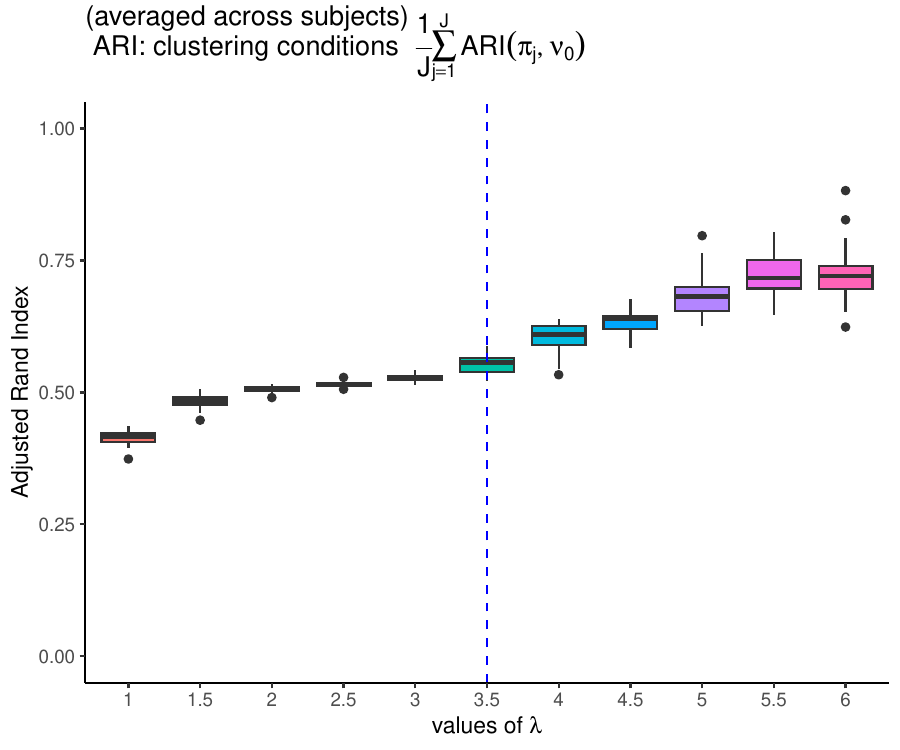}
\end{subfigure}
\begin{subfigure}
\centering
\includegraphics[width=0.35\textwidth]{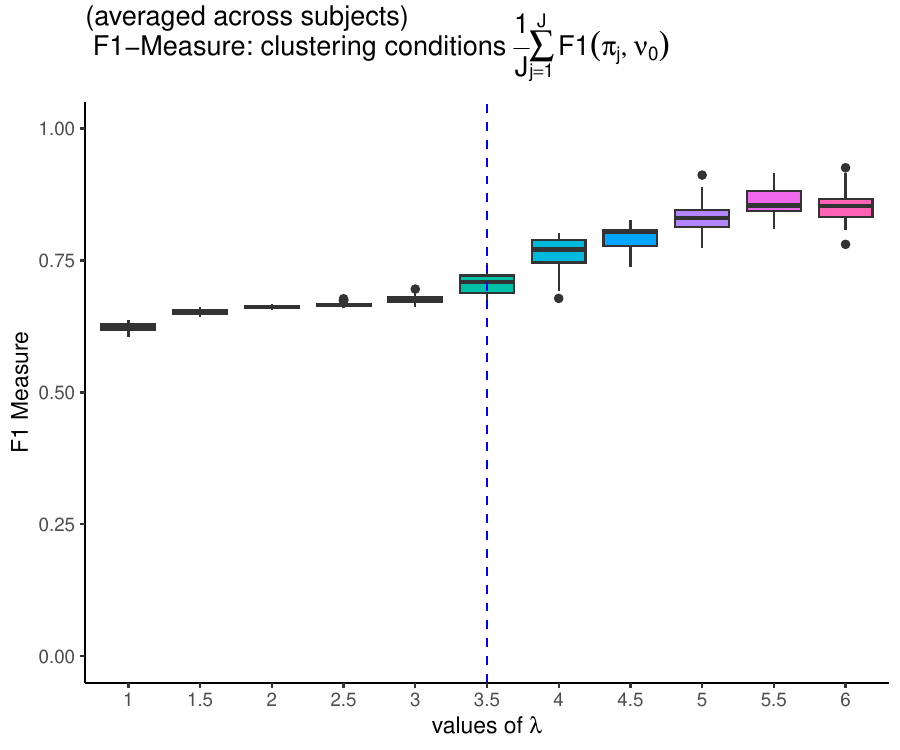}
\end{subfigure}
\\
\begin{subfigure}
\centering
\includegraphics[width=0.35\textwidth]{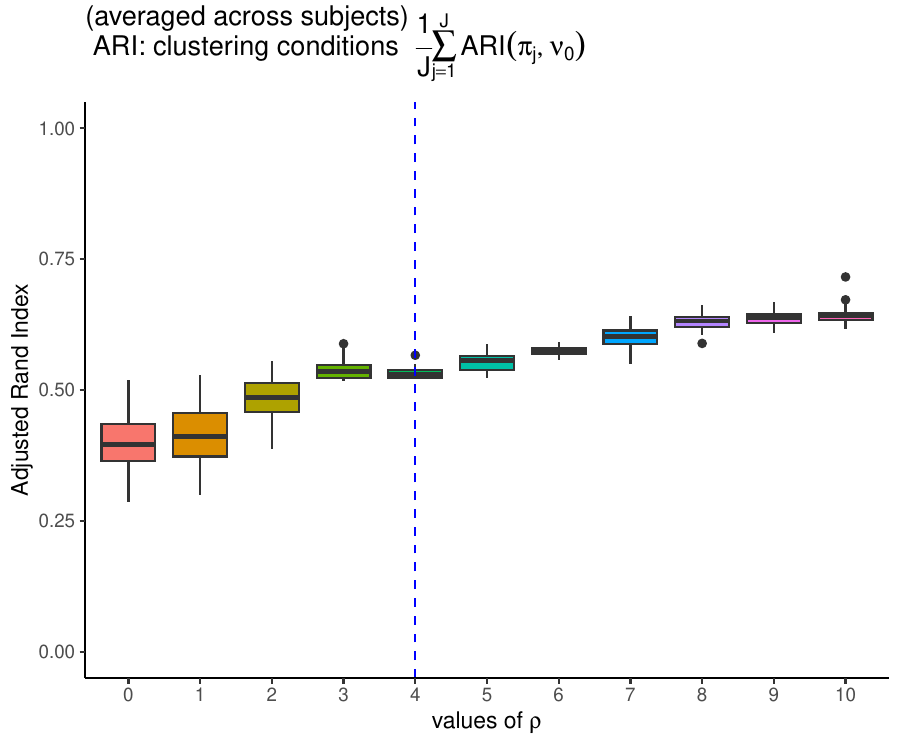}
\end{subfigure}%
\begin{subfigure}
\centering
\includegraphics[width=0.35\textwidth]{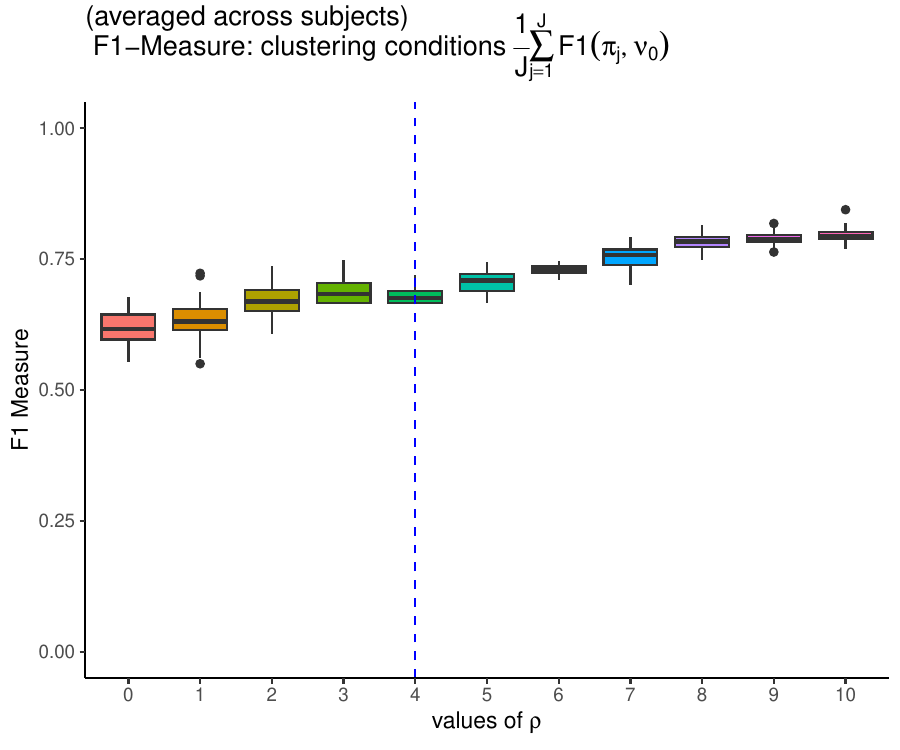}
\end{subfigure}
\caption{\emph{Simulation 2: Sensitivity Analysis.} Average ARI and F1 measures across subjects and 50 replicates, plotted over a grid of $\boldsymbol{\rho}$ and $\boldsymbol{\lambda}$ values (with fixed $\boldsymbol{\tau}=0$). Blue vertical dashed lines mark the selected values.} 
\label{fig:Reviewer1Point7Simulation2_selection}
\end{figure}

In more detail, for \emph{Simulation 2}, our objective is to investigate the ability of the proposed HSP model to cluster conditions within subjects and to compare this performance with that of the HHDP model. To this end, we have included a base partition of conditions, denoted as $\boldsymbol{\nu}_0$, in order to highlight the advantages of our proposed HSP model. The parameter $\boldsymbol{\rho}$ is then pivotal in determining the extent to which one relies on this base partition $\boldsymbol{\nu}_0$. As we are primarily interested in the effect of the base partition of the conditions, we set again $\boldsymbol{\tau}=0$. Figure~\ref{fig:Reviewer1Point7Simulation2_selection} illustrates the sensitivity of the clustering to several values of $\boldsymbol{\rho}$ and $\boldsymbol{\lambda}$, following the previously outlined procedure. We opted for $\boldsymbol{\lambda}=3.5$ and $\boldsymbol{\rho}=4$ since these values strike a favorable balance, offering a good level of similarity to the base partitions. This balance includes some degree of information from the base partition without being overly influenced, with both ARI and F1 measures close to 0.5.   

\begin{figure}[H]
\centering
\begin{subfigure}
\centering
\includegraphics[width=0.4\textwidth]{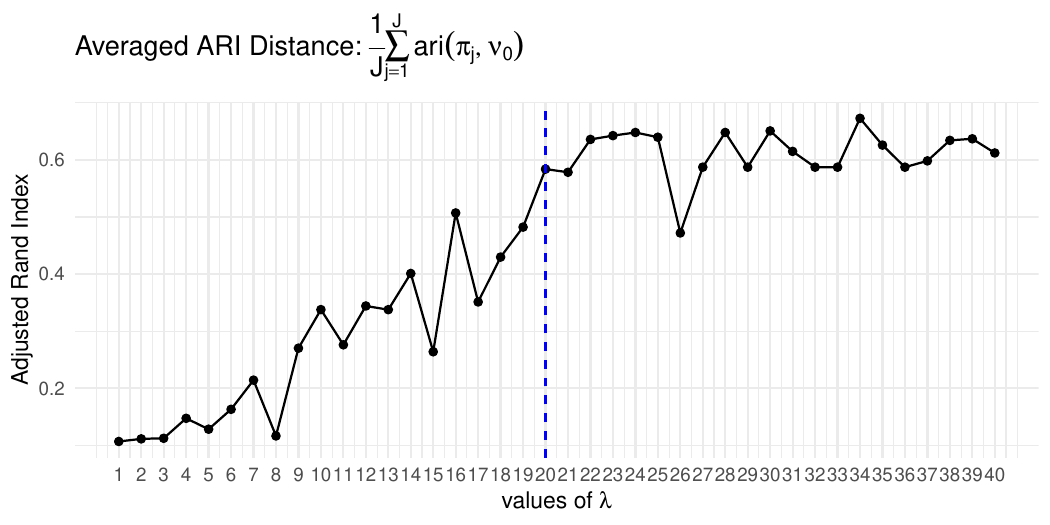}
\end{subfigure}
\begin{subfigure}
\centering
\includegraphics[width=0.4\textwidth]{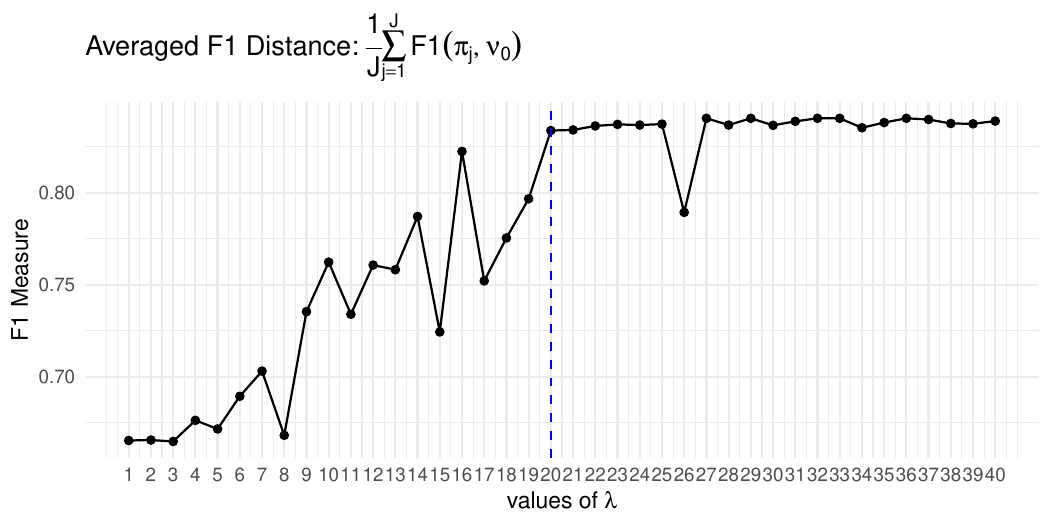}
\end{subfigure}
\\
\begin{subfigure}
\centering
\includegraphics[width=0.4\textwidth]{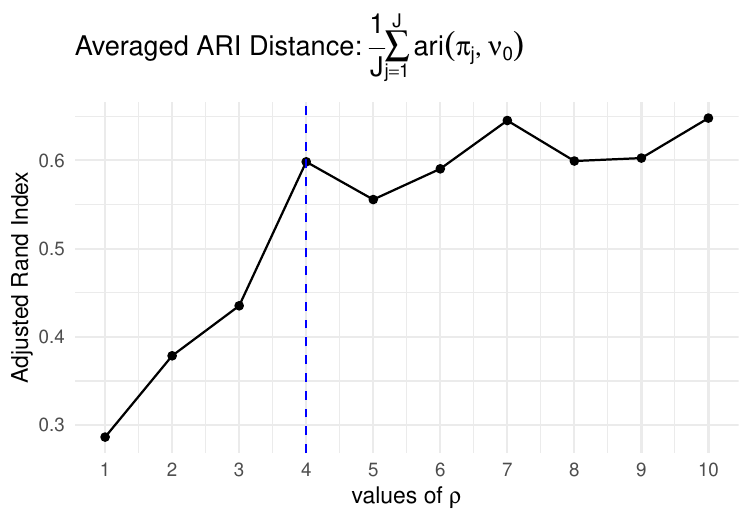}
\end{subfigure}
\begin{subfigure}
\centering
\includegraphics[width=0.4\textwidth]{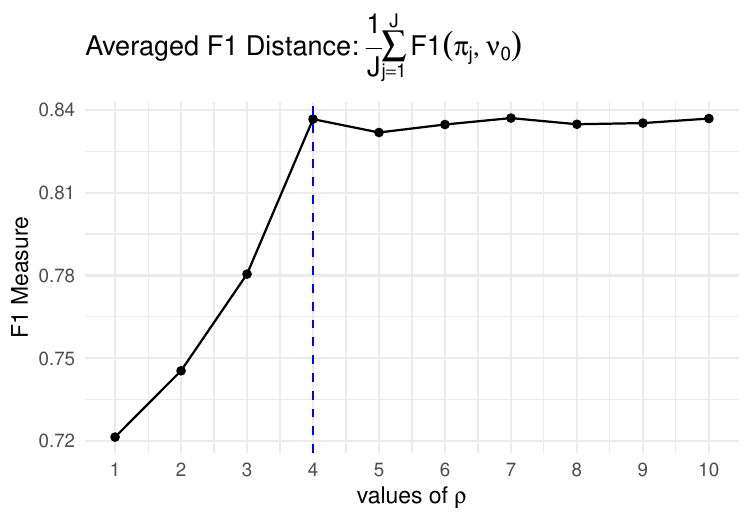}
\end{subfigure}
\\
\begin{subfigure}
\centering
\includegraphics[width=0.4\textwidth]{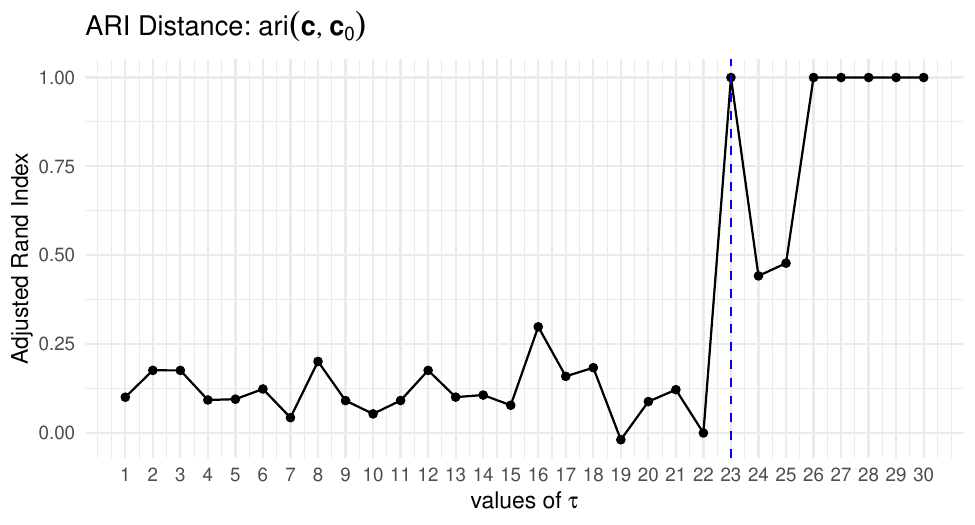}
\end{subfigure}
\begin{subfigure}
\centering
\includegraphics[width=0.4\textwidth]{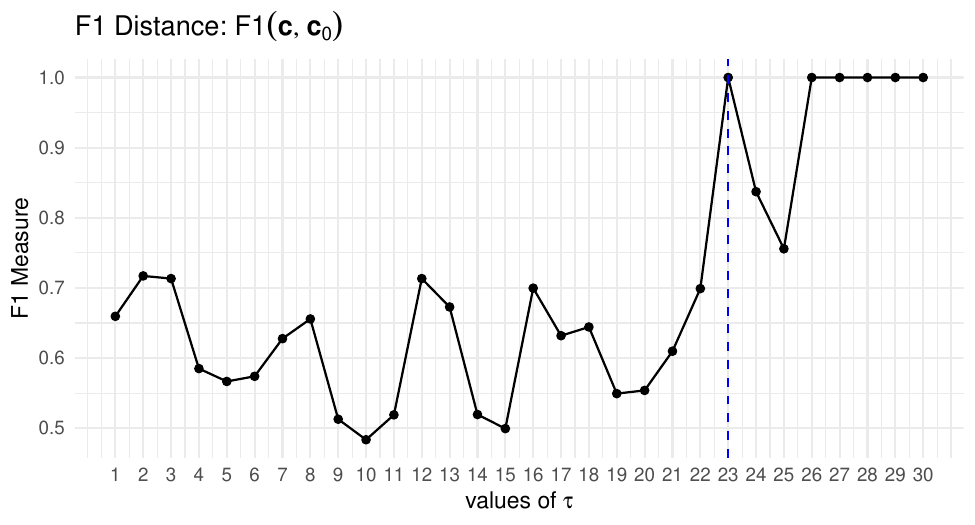}
\end{subfigure}
\caption{\emph{Analysis of mouse-tracking data} The first row displays the average ARI and F1 measures as $\boldsymbol{\lambda}$ varies over a grid of values, with $\boldsymbol{\tau} = 0$ and $\boldsymbol{\rho} \sim Unif\{1, \ldots, 10\}$ (for this plot, $\rho=5$). The second row shows the average ARI and F1 measures as $\boldsymbol{\rho}$ changes over a grid of values, while $\boldsymbol{\tau} = 0$ and $\boldsymbol{\lambda} = 20$. The third row illustrates the ARI and F1 measures as $\boldsymbol{\tau}$ fluctuates over a grid of values, with $\boldsymbol{\rho} = 4$ and $\boldsymbol{\lambda} = 20$. Blue dashed lines in each row mark the selection of the chosen parameter values.}
\label{Fig: R1P7_realdata_select_parameters}
\end{figure}

In the \emph{computer mouse-tracking experiment}, neuroscientists aim to explore how well subject groups from the ERP experiment align with those from mouse tracking. Therefore,  $\boldsymbol{\tau}$ must be a positive value. Following the outlined procedure, in step (c), we draw $\boldsymbol{\rho}$ from a discrete uniform distribution between 1 and 10. Based on the result, we assess the behavior of $\boldsymbol{\lambda}$ over a range of values. As shown in the first row of Figure~\ref{Fig: R1P7_realdata_select_parameters}, the average ARI or F1 initially rises quickly with increases in $\boldsymbol{\lambda}$, and then stabilizes after $\boldsymbol{\lambda}$ surpasses approximately 20. Based on this, we select $\boldsymbol{\lambda}=20$ and proceed to step (e). From the results in the second row of Figure~\ref{Fig: R1P7_realdata_select_parameters}, we set $\boldsymbol{\rho}=4$, as both the ARI and F1 measures appear somewhat stable beyond this point. These values imply that the nested partition of conditions within each subject group is distinct from the base partition of conditions at the population level. Finally, we determine $\boldsymbol{\tau}$ using the chosen values of $\boldsymbol{\rho}$ and $\boldsymbol{\lambda}$ (step h). We evaluate the ARI and F1 between the estimated subject partition $\boldsymbol{c}$ and the base subject partition $\boldsymbol{c}_0$ over a range of $\boldsymbol{\tau}$ values from 1 to 30. As indicated in the third row of Figure~\ref{Fig: R1P7_realdata_select_parameters}, the ARI and F1 measures peak at $\boldsymbol{\tau} = 23$, leading us to select $\boldsymbol{\tau} = 23$ for this case.

\begin{figure}[H]
\centering
\begin{minipage}{.4\textwidth}
\centering
\includegraphics[width=\textwidth]{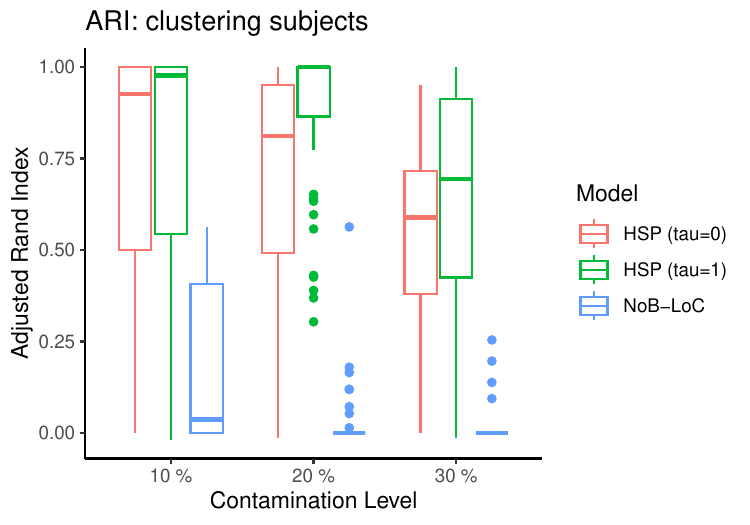}
\end{minipage}%
\hfill 
\begin{minipage}{.4\textwidth}
\centering
\includegraphics[width=\textwidth]{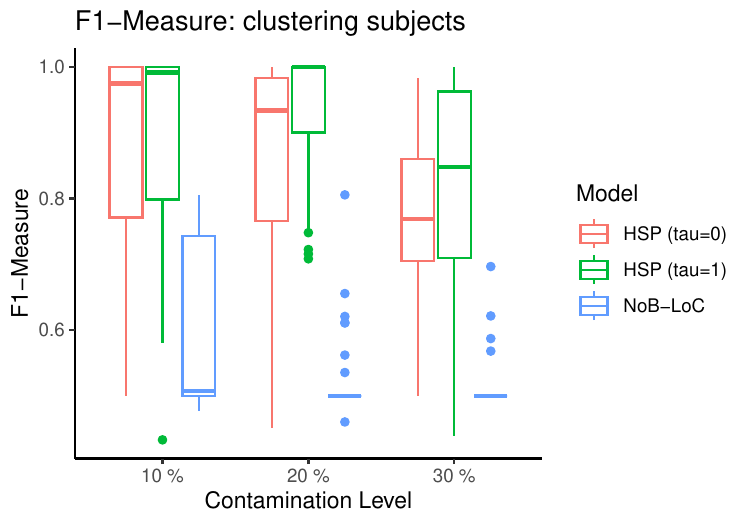}
\end{minipage}
    \vspace{1em} 
\begin{minipage}{.4\textwidth}
\centering
\includegraphics[width=\textwidth]{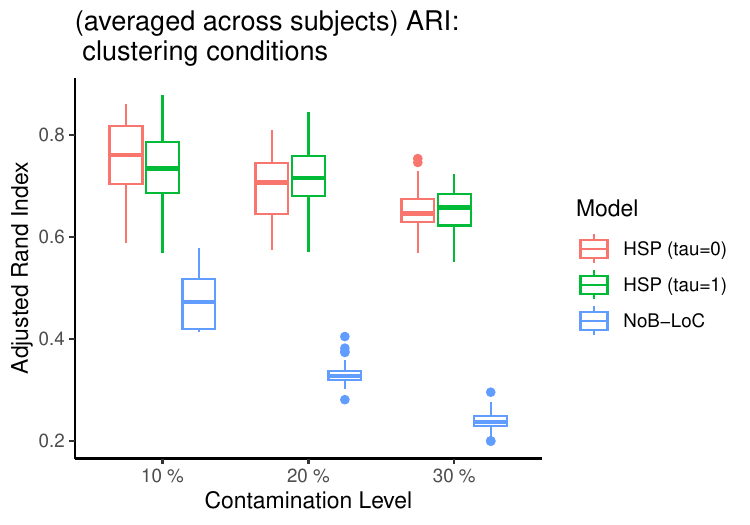}
\end{minipage}%
\hfill
\begin{minipage}{.4\textwidth}
\centering
\includegraphics[width=\textwidth]{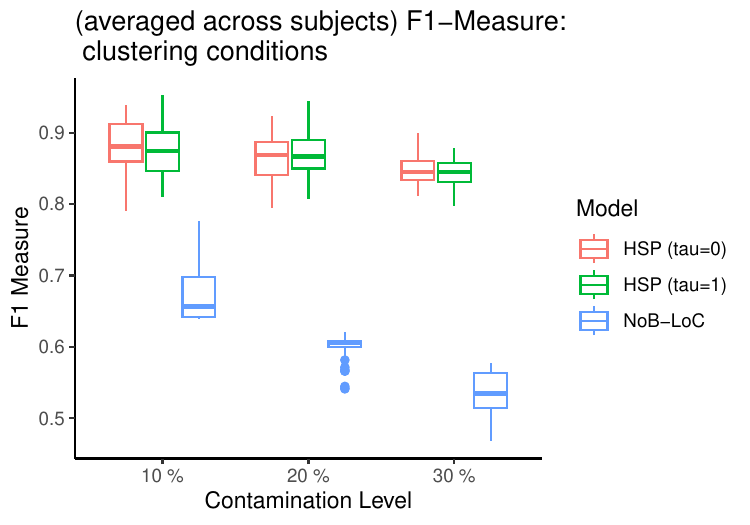}
\end{minipage}
    \caption{{\bf Illustrating the Impact of $\boldsymbol{\tau}>0$ in \emph{Simulation 1(b)}}. Comparison between NoB-LoC and HSP across 50 replications. HSP results are for $\boldsymbol{\tau} = 0, \boldsymbol{\rho} = 0, \boldsymbol{\lambda} = 3.5$, and for $\boldsymbol{\tau}=1, \boldsymbol{\rho}=0, \boldsymbol{\lambda}=3.5$. The first row displays comparisons for the partitioning of 60 subjects, and the second row for the 30 conditions. It also shows the performance of NoB-LoC and HSP under increasing contamination levels (10\%, 20\%, 30\%) within subject clusters}
    \label{fig: R1P7_tau_not_zero}
\end{figure}

{\bf Effect of $\boldsymbol{\tau}>0$ in the simulation setup}. When it is necessary to leverage base partitions of subjects or conditions, $\boldsymbol{\tau}$ and $\boldsymbol{\rho}$ should not be set to zero. In Simulation 1, we deliberately set both $\boldsymbol{\tau}$ and $\boldsymbol{\rho}$ to zero. In Simulation 2, we've set $\boldsymbol{\rho}$ to a non-zero value, demonstrating its role in clustering conditions within subjects. We chose these settings to show that our HSP model is effective even without relying on base partitions and to ensure a fair comparison with other methods. Indeed, using a non-zero $\boldsymbol{\tau}$ leads to improved performance of the HSP model, especially for higher levels of contamination in the setup of \emph{Simulation 1(b)}. Figure~\ref{fig: R1P7_tau_not_zero} compares the HSP's performance with $\boldsymbol{\tau}$ set at 0, as in the main manuscript, and at 1. The figure shows that using a non-zero $\boldsymbol{\tau}$ significantly improves subject clustering. This improvement occurs by including information from the base subject partition and is particularly noticeable in simulations with contaminated data. We illustrate with  $\boldsymbol{\tau} = 1$ for demonstration, but other positive values are also viable to illustrate the key point that $\boldsymbol{\tau}$ may help clustering when leveraging base partitions of subjects.

\section{Large sample size performance}

\begin{figure}[H]
\centering
\begin{minipage}{.49\textwidth}
\centering
\includegraphics[width=\textwidth]{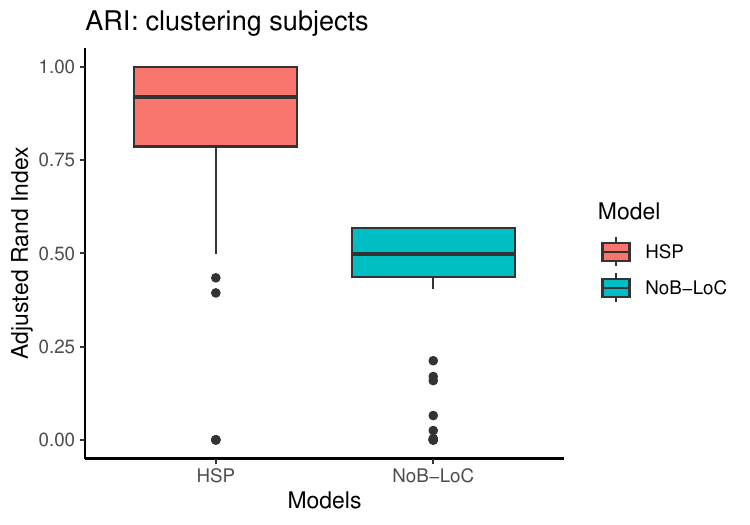}
\end{minipage}%
\hfill 
\begin{minipage}{.49\textwidth}
\centering
\includegraphics[width=\textwidth]{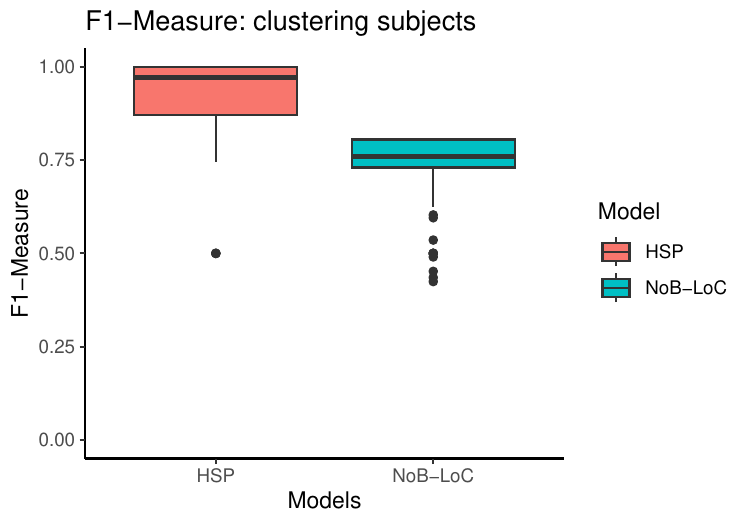}
\end{minipage}
    \vspace{1em} 
\begin{minipage}{.49\textwidth}
\centering
\includegraphics[width=\textwidth]{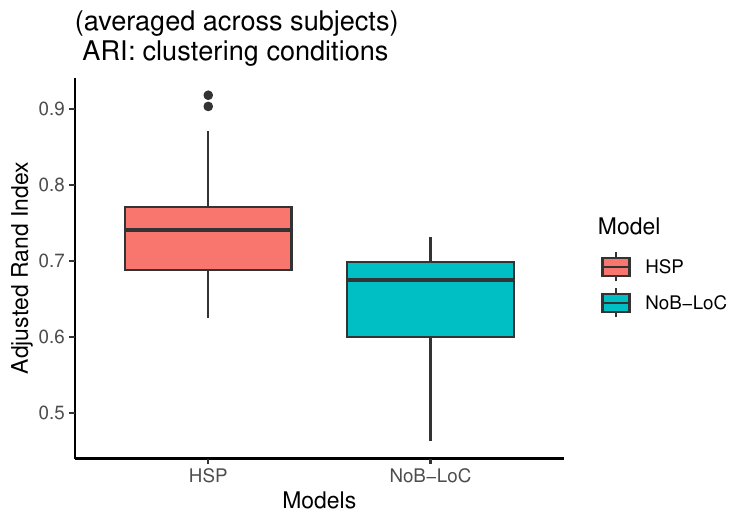}
\end{minipage}%
\hfill
\begin{minipage}{.49\textwidth}
\centering
\includegraphics[width=\textwidth]{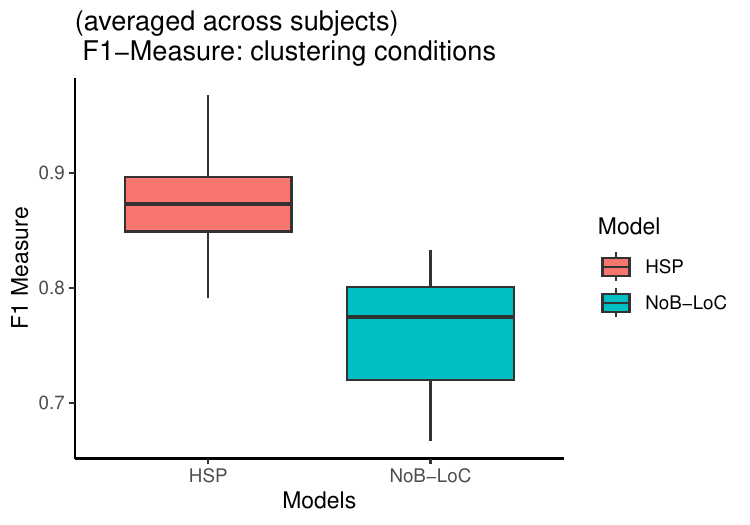}
\end{minipage}
    \caption{A comparison between NoB-LoC and HSP on 50 replications, with 30 conditions and 180 subjects, simulated in the same structure as \emph{Simulation 1(a)}. The results of the HSP shown here are obtained under $\boldsymbol{\tau} = \boldsymbol{\rho} = 0$ and $\boldsymbol{\lambda} = 3.5$, that is without leveraging any base partitions of subjects or conditions provided by prior knowledge. The locations $\mu_{\ell,j}^{*}$ of the condition clusters are generated by permuting $(-0.97, 0.15, 1.37)$ independently across all the subjects. The first line shows the comparisons for the partition of the 180 subjects, while the second line shows the comparisons for the partition of the 30 conditions.}
\label{fig: Reviewer1Point5}
\end{figure}

We conduct a simulation involving a larger number of subjects, $J$, than in the main manuscript. The number of conditions - instead - is typically limited in applications. More specifically, we consider a simulation with  $I = 30$ and $J = 180$,. On our computing server (Intel Xeon Gold 6240R, 2.40 GHz, 192GB RAM), completing 10,000 iterations with 50 replications took under 10 hours. We consider the same true partitions as in \emph{Simulation 1(a)}, with $J=180$ subjects equally divided into three groups. As shown in Figure~\ref{fig: Reviewer1Point5}, the proposed HSP still outperforms the alternative method, NoB-LoC.

\section{Comparison between HSP and BCPlaid}

\begin{figure}[H]
\centering
\begin{minipage}{.6\textwidth}
\centering
\includegraphics[width=0.5\textwidth]{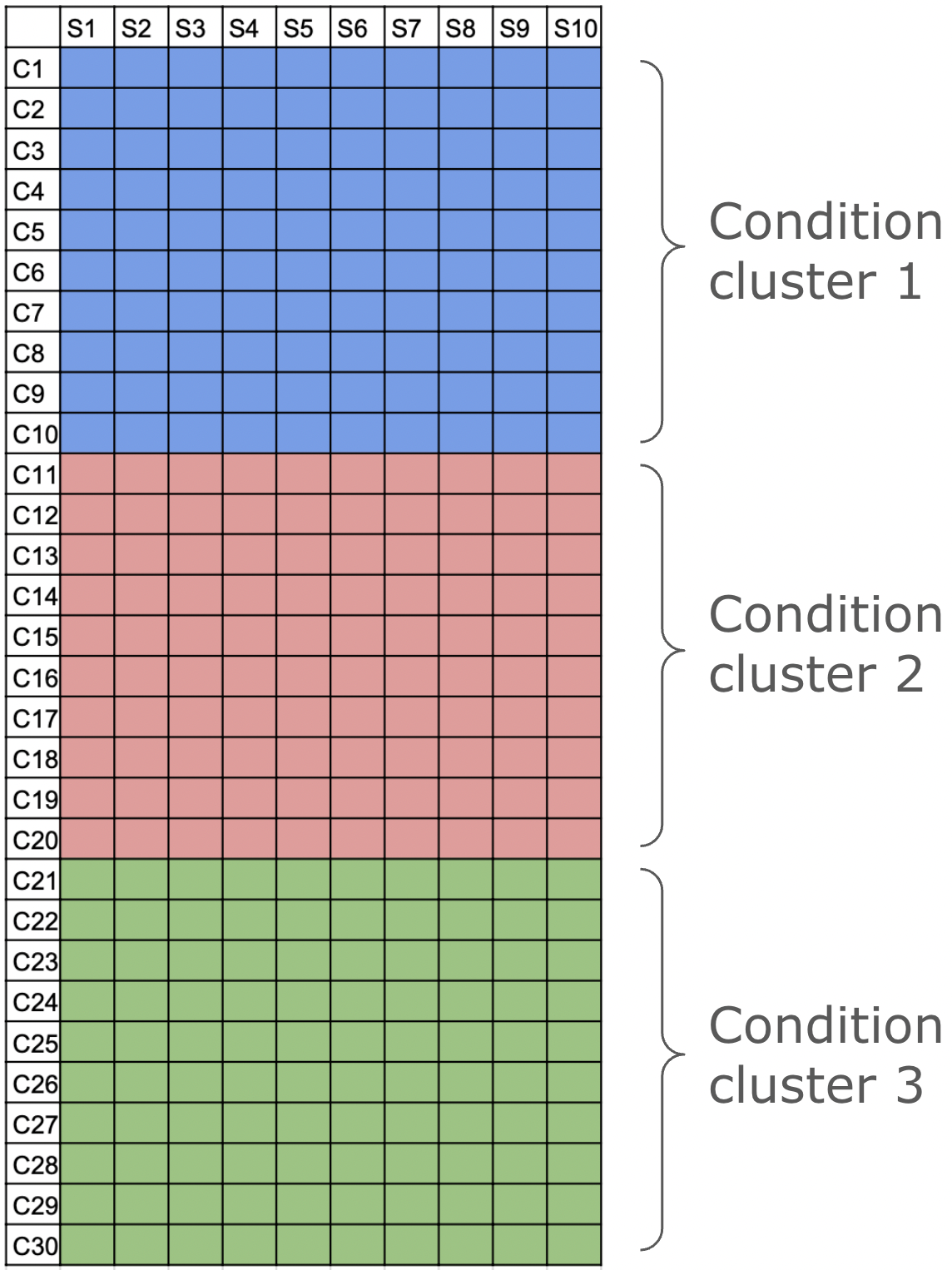}
\end{minipage}%
\caption{The illustration represents a simulation scenario involving 10 subjects and 30 conditions. Conditions are color-coded to indicate clustering based on a shared parameter. The conditions are nested within subjects and partitioned into three distinct clusters. This partitioning is consistent across all subjects.}
\label{fig:R1P4_scenario_plot}
\end{figure}

\begin{figure}[H]
\centering
\begin{minipage}{.6\textwidth}
\centering
\includegraphics[width=\textwidth]{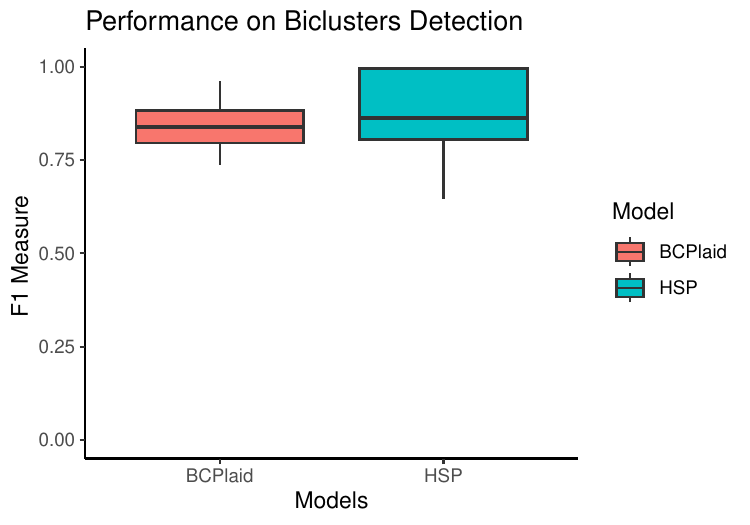}
\end{minipage}%
\caption{{\bf Comparison between HSP and BCPlaid on 50 replicates.} The results of the HSP shown here are obtained under $\boldsymbol{\tau} = \boldsymbol{\rho} = 0$ and $\boldsymbol{\lambda} = 3.5$. It shows the comparison for the bi-clustering of the 30 conditions within the subjects. }
\label{fig:R1P4_compare_BCPlaid}
\end{figure}

We evaluate our model's performance against the improved Plaid model (BCPlaid) as presented by \citep[BCPlaid,][]{turner2005improved} and implemented in the R package \textrm{biclust} \citep{kaiser2008toolbox}. 
As a fair comparison with the BCPlaid model, the comparison was conducted under a scenario where a group of subjects shares the same nested partition of conditions. We simulated data across $J=10$ subjects under $I=30$ conditions, with the subjects sharing a nested partition of conditions. The data points within each subject were partitioned with clustering labels $\{\boldsymbol{1_{10}, 2_{10}, 3_{10}}\}$ (e.g., $\boldsymbol{1_{10}} = \{1,1,1,1,1,1,1,1,1,1\}$). Reflecting the patterns found in the actual data, we generated $y_{ij} \sim N(\mu_{\ell}^{*}, \sigma_{\ell}^{2 *})$ with $\sigma_{\ell}^{2 *} = 0.16$ and $(\mu_{1}^{*}, \mu_{2}^{*}, \mu_{3}^{*}) = (-0.97, 0.15, 1.37)$. Therefore, we confined our model to scenarios where a shared parameter defines the bicluster. Figure~\ref{fig:R1P4_scenario_plot} presents a stylized data matrix organized according to the cluster membership of conditions, as determined from the simulated dataset. This arrangement illustrates a scenario where a group of subjects shares an identical nested partition of the conditions.  As shown in Figure~\ref{fig:R1P4_compare_BCPlaid}, in situations where a group of subjects shares a common nested partition of conditions, our HSP model demonstrates superior performance over BCPlaid in the bi-clustering of these conditions.  

This comparison highlights the distinct characteristics that set our HSP model apart from other biclustering methods. The HSP avoids overlapping in its clustering configurations and does not require a common parameter within submatrices. Nevertheless, it effectively captures the clustering allocations,  when a group of subjects, in reality, shares a common nested partition of conditions.

\end{document}